\documentclass[traditabstract]{aa}
\usepackage{graphicx}

\usepackage{epsfig}
\usepackage[colorlinks=true,linkcolor=blue]{hyperref}

\newcommand{\be}{\begin{equation}}
\newcommand{\ee}{\end{equation}}

\newcommand{\bea}{\begin{eqnarray}}
\newcommand{\eea}{\end{eqnarray}}

\newcommand{\al}{\alpha}

\newcommand{\gm}{\gamma}
\newcommand{\Gm}{\Gamma}
\newcommand{\dl}{\delta}
\newcommand{\Dl}{\Delta}
\newcommand{\eps}{\epsilon}

\newcommand{\Lm}{\Lambda}

\newcommand{\rh}{\rho}

\newcommand{\sg}{\sigma}

\newcommand{\ch}{\chi}

\newcommand{\om}{\omega}
\newcommand{\Om}{\Omega}

\newcommand{\rarrow}{\rightarrow}
\newcommand{\Rarrow}{\Rightarrow}

\newcommand{\nn}{\nonumber}

\newcommand{\varep}{\varepsilon}

\begin{document}

\title{Polytropic dark matter flows illuminate dark energy and accelerated expansion}

\author{K. Kleidis \inst{1}
        \and
        N. K. Spyrou \inst{2}}

\institute{Department of Mechanical Engineering, Technological Education Institute of Central Macedonia, 621.24 Serres, Greece (Hellas). \\e-mail: \texttt{kleidis@teiser.gr} \and Department of Astronomy, Aristoteleion University of Thessaloniki, 541.24 Thessaloniki, Greece (Hellas). \\e-mail: \texttt{spyrou@auth.gr}}

\offprints{K. Kleidis}

\date{Received ..... ; accepted .....}

\titlerunning{Polytropic DM flows illuminate DE and ...}

\authorrunning{K. Kleidis \& N. K. Spyrou}

\abstract{Currently, a large amount of data implies that the matter constituents of the cosmological dark sector might be collisional. An attractive feature of such a possibility is that, it can reconcile dark matter (DM) and dark energy (DE) in terms of a single component, accomodated in the context of a polytropic-DM fluid. In fact, polytropic processes in a DM fluid have been most successfully used in modeling dark galactic haloes, thus improving significantly the velocity dispersion profiles of galaxies. Motivated by such results, we explore the time evolution and the dynamical characteristics of a spatially-flat cosmological (toy-)model, in which, in principle, there is no DE at all. Instead, in this model, the DM itself possesses some sort of fluid-like properties, i.e., the fundamental units of the Universe matter-energy content are the volume elements of a DM fluid, performing polytropic flows. In this case, together with all the other physical characteristics, the (thermodynamic) energy of this fluid's internal motions is (also) taken into account as a source of the universal gravitational field. This (conventional) form of energy can compensate for the extra (dark) energy, needed to compromise spatial flatness, namely, to justify that, today, the total-energy density parameter is exactly unity. The polytropic cosmological model, depends on only one free parameter, the corresponding (polytropic) exponent, $\Gm$. What makes this model particularly interesting, is that, for $\Gm \leq 0.541$ (and without the need for either any exotic DE or the cosmological constant), the (conventional) pressure becomes negative enough, so that the Universe accelerates its expansion at cosmological redshifts below a transition value. In fact, several physical reasons, e.g., the cosmological requirement for cold DM (CDM) and a positive velocity-of-sound square, impose further constraints on the value of $\Gm$, which, eventually, is settled down to the range $-0.089 < \Gm \leq 0$. Such a cosmological model does not suffer either from the age problem or from the coincidence problem. At the same time, this model reproduces (to high accuracy) the distance measurements performed with the aid of the supernovae (SNe) Type Ia standard candles, and most naturally interprets, not only when, but, also, why the Universe transits from deceleration to acceleration, thus arising as a mighty contestant for a (conventional) DE model.}


%\pacs{04.40.Nr, 95.36.+x, 98.80.Es, 98.80.Jk}

\maketitle

\section{Introduction}

The study of the cosmic microwave background (CMB) has proven to be a powerful tool in exploring the post-recombination Universe. In particular, observations of the temperature variations in the CMB have provided strong evidence that the Universe emerged out of the radiation epoch as a spatially-flat Robertson-Walker (RW) model (see, e.g., de Bernardis et al. 2000; Jaffe et al. 2001; Padin et al. 2001; Stompor et al. 2001; Netterfield et al. 2002; Spergel et al. 2003, 2007; Komatsu et al. 2009, 2011). 

This result, however, implies that, the overall energy density, $\varep$, of the Universe matter-energy content, in units of the quantity $\varep_c = \rh_c c^2$ (equivalent to the critical rest-mass density, $\rh_c= \frac{3 H_0^2}{8 \pi G}$, where $H_0$ is the Hubble parameter at the present epoch, $c$ is the velocity of light, and $G$ is Newton's universal constant of gravitation), should be very close to unity, $\Om = \frac{\varep}{ \varep_c} \simeq 1$, i.e., much larger than the (most recenly) measured value of the density parameter, $\Om_M = \frac{\rh}{\rh_c} = 0.279$ (Hinshaw et al. 2013). In other words, (either) an extra amount of energy (or a component that has not, yet, been taken into account) is needed, to compromise spatial flatness.

The need for a distributed extra-energy component, i.e., one that is not associated with concentrations of mass, is suggested also by high-precision distance measurements, performed with the aid of the SNe Ia standard candles (Hamuy et al. 1996; Garnavich et al. 1998; Perlmutter et al. 1998, 1999$a$; Schmidt et al. 1998; Riess et al. 1998, 2001, 2004, 2007;  Knop et al. 2003; Tonry et al. 2003; Barris et al. 2004; Krisciunas et al. 2005; Astier et al. 2006; Jha et al. 2006; Miknaitis et al. 2007; Wood-Vasey et al. 2007; Amanullah ey al. 2008, 2010; Holtzman et al. 2008; Kowalski et al. 2008; Hicken et al. 2009$a$, 2009$b$; Kessler et al. 2009; Contreras et al. 2010; Guy et al. 2010; Suzuki et al. 2012). In view of these measurements, in a cosmological model with matter content in the form of dust (i.e., of vanishing pressure), the far-off light-emitting sources appear to be dimmer (i.e., they seem to lie farther away) than what is theoretically expected. In confronting with this result, Riess et al. (1998) and Perlmutter et al. (1999$a$) admitted that, theory (e.g., Carroll et al. 1992) meets observation only in the case of a non-zero cosmological constant, $\Lm$, so that $\Om_M \simeq 0.3$ and $\Om_{\Lm} = \frac{\Lm c^2}{3 H_0^2} \simeq 0.7$ (see also Riess et al. 2004). But, a non-vanishing cosmological constant is related to a repulsive component of the gravitational force (see, e.g., Sahni 2004) and, therefore, the apparent dimming of the distant light-emitting sources was subsequently attributed to a (relatively recent) phase of accelerated expansion (see, e.g., Frieman et al. 2008; Linder 2008).

Apart from Enstein's "bigest blunder", the particle-physics vacuum does contribute an effective cosmological constant, which could serve, also, as compensation to the extra energy needed to flatten the Universe (Sahni \& Starobinsky 2000). Unfortunately, the energy density attributed to this source is $10^{123}$ times larger than what is observed (see, e.g., Padmanabhan 2003), hence, an approach other than the cosmological constant could (also) be useful. 

In this context, many physically-motivated models have appeared in the literature, including quintessence (Caldwell et al. 1998), K-essence (Armendariz-Picon et al. 2001), phantom Cosmology (Caldwell 2002) and tachyonic matter (Padmanabhan 2002), involving (also) several braneworld scenarios, such as DGP-gravity (Dvali et al. 2000) and the landscape scenario (Bousso \& Polchinski 2000), as well as alternative-gravity theories, such as the scalar-tensor theories (Esposito-Farese \& Polarski 2001) and $f(R)$-gravity (Capozziello et al. 2003), holographic gravity (Cohen et al. 1999; Li 2004; Pav\'{o}n \& Zimdahl 2005), Chaplygin gas (Kamenshchik et al. 2001; Bento et al. 2002; Bean \& Dor\'{e} 2003; Sen \& Scherrer 2005), Cardassian cosmology (Freese \& Lewis 2002; Gondolo \& Freese 2003; Wang et al. 2003), theories of compactified internal dimensions (Mongan 2001; Defayet et al. 2002; Perivolaropoulos 2003; Sami et al. 2004), mass-varying neutrinos (Fardon et al. 2004; Peccei 2005), and so on (for a detailed review see, e.g., Caldwell \& Kamionkowski 2009).

In the meantime, the list of the observational data in favor of a distributed extra-energy component continued to grow, including evidence from galaxy clusters (Allen et al. 2004), the integrated Sachs-Wolfe (ISW) effect (Boughn \& Crittenden 2004), baryon acoustic oscillations (BAOs) (Eisenstein et al. 2005; Percival et al. 2010), weak gravitational lensing (WGL) (Huterer 2002; Copeland et al. 2006), and the Lyman-$\al$ (LYA) forest (Seljak et al. 2006). A combination of these data with those from the Wilkinson microwave anisotropy probe (WMAP) survey (see, e.g., Dunkley et al. 2009) has provided evidence for cosmic acceleration at the $5 \sg$ level. This could no longer be ignored, and the 2011 Nobel Prize in Physics came to officially validate it (in connection, see Riess 2012).

Eventually, both the extra energy (needed to compromise spatial flatness) and the accelerated expansion of the Universe (implemented to justify the unexpected dimming of the SNe Ia standard candles) were reconciled by another assumption, that of an exotic fluid endowed with negative pressure, which has been termed dark energy (Turner \& White 1997; Perlmutter et al. 1999$b$), reflecting our ignorance on its exact nature (for a review of the various DE models see, e.g., Miao et al. 2011). However, a dark component (namely, the DM itself) was already present in the energy density of the Universe matter content. 

Indeed, it is rather well-established that, almost $84 \%$ (by mass) of the matter in the Universe consists of non-luminous and non-baryonic material (see, e.g., Tegmark et al. 2006; Hinshaw et al. 2013). In fact, recent observations based on WGL, suggest that the matter distribution of galaxies extends beyond the virial radius, roughly to the middle of the neighbouring galaxies (Masaki et al. 2012, in connection, see also Spyrou 2011). Such a mass distribution could explain the gap, observed in the mass-density profiles, between the global value $\Om_M = 0.279$ and the typical value $\Om_{gal} = 0.150$, arising from the luminosity density of the galaxies multiplied by the corresponding mass-to-light ratio. 

Although we do not know for certain how the DM came to be formed, a sizeable relic abundance of weakly-interacting massive particles (WIMPs) is generally expected to have been produced as a by-product of the Universe's hot youth (see, e.g., Kolb \& Turner 1990, p. 369). These particles decouple from radiation much earlier than pure-baryonic matter does. Hence, very soon after recombination $(t_R)$, the baryons fall into deep potential wells of the already evolved DM-perturbations and become bounded to them, i.e., for $t > t_R$, there are no freely-floating baryons around (see, e.g., Olive 2003, Hooper 2009). 

As far as structure formation is concerned, all forms of DM are not equivalent. Particles which are still highly relativistic at $t_R$ (like neutrinos or other particles with masses lower than $100 \: eV/c^2$) have the property that, due to free streaming, they erase perturbations out to very large scales (Bond et al. 1980). It is, therefore, expected that, very-large-scale structures form first, and fragmented, to form galaxies, later. Particles with this property are termed hot dark matter (HDM). On the other hand, CDM (i.e., particles with masses larger than $1 \: MeV/c^2$) has the opposite behavior: Small-scale structures form first, aggregating to form larger structures later (Bond \& Szalay 1983). It is now well-known that pure HDM cosmologies can not reproduce the observed large-scale structure of the Universe (see, e.g., Klypin et al. 1993). On the contrary, CDM models are in remarkably-good agreement with the observed power-spectrum of LYA absorbers (Croft et al. 1999). Apart from debating on the precise nature of the DM constituents, the scientific community used to argue that they should be collisionless. 

However, many results from the high-energy particle detector PAMELA (Adriani et al. 2009), combined with data from the WMAP survey (Hooper et al. 2007), have revealed an unusually high electron - positron production in the Universe, much more than what is anticipated by SNe explosions or cosmic-ray collisions. These results have led many scientists to argue that among the best candidate sources of these high-energy events are the annihilations of WIMPs (see, e.g., Barger et al. 2008; Bergstrom et al. 2008; Cirelli \& Strumia 2008; Regis \& Ullio 2008; Baushev 2009; Cholis et al. 2009a, 2009b; Fornasa et al. 2009; Fox \& Poppitz 2009; Kane et al. 2009; Zurek 2009), i.e., that the DM constituents can be slightly collisional (Spergel \& Steinhardt 2000; Arkani-Hamed et al. 2009; Cirelli et al. 2009; Cohen \& Zurek 2010; Van den Aarssen et al. 2012), although, there are studies that disagree with this interpretation (see, e.g., Feng et al. 2010). An attractive feature of a collisional-DM model is that, it can describe both DM and DE in terms of a single component. In other words, a cosmological model filled with self-interacting DM could be a relatively inexpensive solution to the DE problem. Accordingly, several ways of accommodating both the DM and the DE into a unified theoretical framework have been considered (see, e.g., Zimdahl et al. 2001; Bili\'{c} et al. 2002; Balakin et al. 2003; Gondolo \& Freese 2003; Makler et al. 2003; Scherrer 2004; Ren \& Meng 2006; Meng et al. 2007; Lima et al. 2008, 2010, 2012; Basilakos \& Plionis 2009, 2010; Dutta \& Scherrer 2010; Xu et al. 2012), not always without been disputed (see, e.g., Sandvik et al. 2004).

In this context, in a recent work by Kleidis \& Spyrou (2011), it was suggested that the self-interacting DM could, phenomenologically, attribute to the Universe matter-content some sort of fluid-like properties, and (so) lead to a conventional approach to the DE concept. Indeed, the main problem of the current cosmological picture is that, the Universe must contain an amount of energy, which is considerably larger than that equivalent to the total rest-mass of its matter content. However, if the DM constituents collided with each other frequently enough, enabling their (kinetic) energy to be re-distributed, i.e., if the DM itself possessed some sort of thermodynamic properties, a conventional extra-energy component might be present in the Universe, given by the energy of the internal motions of the collisional-DM fluid. 

The difference between such a model and those of the classical Friedmann-Robertson-Walker (FRW) cosmology used so far, is that, in this case, the fundamental units of the Universe matter content are the volume elements of the collisional-DM fluid. Hence, together with all the other physical characteristics, the energy of this fluid's internal motions is also taken into account as a source of the universal gravitational field, thus resulting in a self-consistent, conventional approach to the DE concept. Nevertheless, according to Kleidis \& Spyrou (2011), a model in which the DE is attributed to the isothermal flows of a collisional-DM fluid, although it is intriguing, it has two delicate points: Not only it is decelerating, but, also, it is incompatible to the observational data currently available, unless the matter content of the dark sector consists of HDM. 

In an effort to confront with these issues, we note that, in a realistic cosmology, the polytropic motion of the cosmic fluid's volume elements is much more physically relevant than the corresponding isothermal flow (see, e.g., Christensen-Dalsgard 2008, pp. 64-69). In particular, polytropic processes in a DM fluid have been most successfully used in modeling dark galactic haloes, thus improving significantly the velocity dispersion profiles of galaxies (Bharadwaj \& Kar 2003; Nunez et al. 2006; Zavala et al. 2006; B{\"o}hmer \& Harko 2007; Saxton \& Wu 2008; Su \& Chen 2009; Saxton \& Ferreras 2010). Clearly, we can not help but wondering, what would be the impact of such an assumption on cosmological scale.

Polytropic (DM) cosmological models were first encountered as natural candidates for Cardassian Cosmology models (see, e.g., Freese \& Lewis 2002; Gondolo \& Freese 2003; Wang et al. 2003; Freese 2005). They have also been used as phenomenological models of DE (see, e.g., Nojiri et al. 2005; Stefanci\'{c} 2005; Mukhopadhyay et al. 2008), especially, in an effort to establish an interaction between the exotic DE fluid and its conventional (DM) counterpart (see, e.g., Karami et al 2009; Karami \& Abdolmaleki 2010$a$, 2010$b$, 2012; Malekjani et al. 2011; Chavanis 2012$a$, 2012$b$, 2012$c$; Karami \& Khaledian 2012; Asadzadeh et al. 2013). 

Our approach, however, is totally different and, in the reasoning of {\em Occam's razor}, rather preferable, since, it does not involve any DE at all. Instead, in the present article, we examine the evolution and the dynamical properties of a cosmological (toy-)model, driven by a gravitating fluid with thermodynamical content, consisting of DM (dominant) and baryonic matter (subdominant). The fundamental matter constituents of this model are the volume elements of the DM fluid, performing polytropic flows. As a consequence, now, the energy of this fluid's internal motions is also taken into account as a source of the universal gravitational field. This (conventional) form of energy can compensate for the extra (dark) energy, needed to compromise spatial flatness, namely, to justify that, today, the total-energy density parameter is exactly unity. The polytropic model depends on only one free parameter, the corresponding (polytropic) exponent, $\Gm$. What makes such a model particularly interesting, is that, for $\Gm \leq 0.541$, the (conventional) pressure becomes negative enough, in the sense that the Universe accelerates its expansion, at cosmological redshifts below a transition value. In fact, several physical reasons, e.g., the cosmological requirement for CDM and a positive velocity-of-sound square, may lead to successive constraints on $\Gm$, the value of which, eventually, is settled down to the range $-0.089 < \Gm \leq 0$. The polytropic-DM cosmological model that we propose, does not suffer either from the age problem or from the coincidence problem. At the same time, this model reproduces (to high accuracy) the distance measurements performed with the aid of the SNe Ia standard candles, without the need for any exotic DE or the cosmological constant. In this context, it is worth noting that, the value of the CMB-shift parameter in the $\Lm$CDM limit of our model reproduces, to high accuracy, the corresponding result obtained by fitting the observational (CMB) data to the standard $\Lm$CDM model. Finally, the polytropic-DM model most natutally interprets not only when, but, also, why the Universe transits from deceleration to acceleration, thus arising as a mighty contestant for a (conventional) DE model. However, we need to stress that, it is not yet clear what kind of micro-physics would give the postulated polytropic behavior; hence, our model is to be seen as an effective (phenomenological) approach of an elementary-physics scenario yet to be discovered (in connection, see, e.g., Gondolo \& Freese 2003; Arkani-Hamed et al. 2009; Van den Aarssen et al. 2012).

This paper is organized as follows: In Sect. 2, we summarize the thermodynamical aspect of polytropic processes in an expanding Universe. Accordingly, in Sect. 3, we explore the dynamical characteristics of a spatially-flat cosmological model in which, in principle, there is no DE at all. The evolution of this model is driven by an ideal fluid, consisting (mainly) of (thermodynamically-involved) DM, the volume elements of which perform polytropic flows. The corresponding results suggest that, in the context of the polytropic treatment, \texttt{(i)} the extra (dark) energy (needed to compromise spatial flatness) can be compensated by the energy of the internal motions of the DM fluid, \texttt{(ii)} the Universe does not suffer either from the age or the coincidence problems, and \texttt{(iii)} such a cosmological model (most naturally) accelerates its expansion at redshifts lower than a transition value, which depends only on the polytropic exponent, $\Gm$. What is more important, is that, as we demonstrate in Sect. 4, in a polytropic (DM) Universe the theoretically-derived distance modulus fits (to high accuracy) the Hubble diagram of an extended sample of SNe Ia standard candles, thus emerging as a mighty contestant for a (conventional) DE model. Finally, in Sect. 5, we give a clear physical interpretation of why and when the Universe transits from deceleration to acceleration, and we conclude in Sect. 6.

\section{Thermodynamics of polytropic processes}

A polytropic change is a reversible (or irreversible) process, along which, the specific heat of a thermodynamical system, \be {\cal C} = \frac{dQ}{dT} \: , \ee varies in a prescribed manner (see, e.g., Horedt 2004, p. 2). An important special case is when ${\cal C}$ is constant. In this case, along with the fundamental equation of state, \be p \propto \rh T \: , \ee which relates pressure $(p)$ to rest-mass density $(\rh)$ and absolute temperature $(T)$ in a perfect fluid, there is also a second equation, namely ${\cal C} = constant$, and the polytropic thermodynamical system is left with only one independent state-variable; in our case, this will be the rest-mass density, i.e., the part, equivalent to the energy density $\rh c^2$, that remains unaffected by the internal motions of the cosmic fluid (barotropic flow). Accordingly, Eq. (2) is decomposed to \be p = p_0 \left ( \frac{\rh}{ \rh_0} \right )^{\Gm} \ee and \be T = T_0 \left ( \frac{\rh}{\rh_0} \right )^{\Gm - 1} \ee  (see, e.g., Chandrasekhar 1939, p. 85; Horedt 2004, p. 9), where $p_0$, $\rh_0$ and $T_0$ denote the present-time values of pressure, rest-mass density and temperature, respectively, and $\Gm$ is the polytropic exponent, defined as \be \Gm = \frac{{\cal C}_P - {\cal C}}{{\cal C}_V - {\cal C}} \: , \ee (see, e.g., Chandrasekhar 1939, p. 86; Horedt 2004, p. 5) where ${\cal C}_P$ (${\cal C}_V$) is the specific heat at constant pressure (volume).

For ${\cal C} = 0$, the polytropic process is reduced to an adiabatic one $(dQ = 0)$, while, for ${\cal C} \rarrow \pm \infty$, it results in an isothermal process $(dT = 0)$. On the other hand, for ${\cal C} = {\cal C}_P$, the polytropic becomes an isobaric process $(p = constant)$, and, for ${\cal C} \rarrow {\cal C}_V$, it yields an isochoric (or isometric) one, although, in an expanding Universe, this kind of process is possible only as a limiting case (static Universe). Clearly, a polytropic change can be considered as a general treatment, including many (different) thermodynamical processes (as well as their transitional phases) in a single formula. Although ${\cal C}$ is the real, physical quantity, it is useful to parametrize it in terms of $\Gm$, and so, address the impact of the polytropic exponent to each and everyone of these processes. Accordingly, we express Eq. (5) in the form \be {\cal C} = \left [1 - \left ( \frac{\gm - 1}{\Gm - 1} \right ) \right ] {\cal C}_V \: , \ee where \be \gm = \frac{{\cal C}_P}{{\cal C}_V} \ee is the adiabatic exponent. Indicative only, for $\gm = \frac{5}{3}$, the behavior of ${\cal C}$ as a function of $\Gm$, is presented in Fig. 1. The various processes encountered in it, are summarized in Table I. 

\begin{figure}[ht!]
\centerline{\mbox {\epsfxsize=9.cm \epsfysize=7.cm
\epsfbox{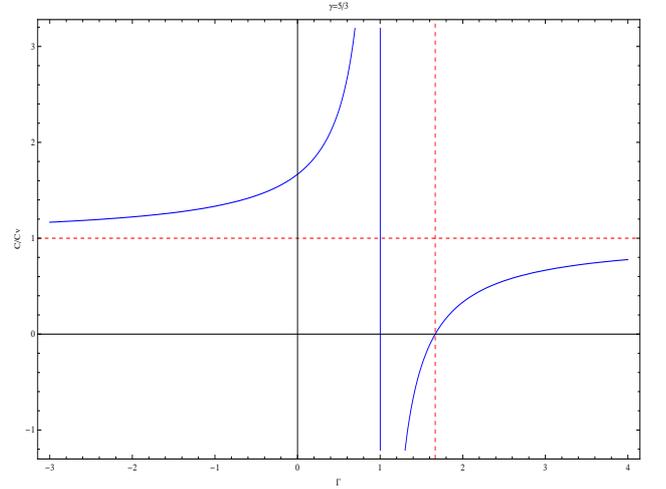}}} \caption{Behavior of the specific heat, ${\cal C}$ (in units of ${\cal C}_V$, as a function of the polytropic exponent, $\Gm$ (blue solid curve), for $\gm = \frac{5}{3}$ (vertical red dashed line).}
\end{figure}

As we observe, the concept of polytropic process is of great importance (at least) from a methodological viewpoint, since it correlates all other known thermodynamical processes. 

\begin{table}[ht!]
\begin{tabular}{|l|l|l|}
  \hline
   {\bf Value of $\Gm$}   & {\bf Value of ${\cal C}$}            & {\bf Process in an expanding Universe}                                 \\
  \hline
   $\Gm \rarrow - \infty$ & ${\cal C} \rarrow {\cal C}_{V_+}$    & \texttt{isochoric limit}                                               \\
  \hline
   $\Gm < 0$              & ${\cal C}_V < {\cal C} < {\cal C}_P$ & \texttt{super-isobaric} (see, e.g., Kamiuto 2008)                      \\
  \hline
   $\Gm = 0$              & ${\cal C} = {\cal C}_P$              & \texttt{isobaric} - equivalent to $\Lm$CDM model (cf. Eq. (64), below) \\
  \hline
   $0 < \Gm < 1$          & ${\cal C} > {\cal C}_P$              & \texttt{sub-isobaric}                                                  \\
  \hline
   $\Gm \rarrow 1_{-}$    & ${\cal C} \rarrow + \infty$          & \texttt{isothermal limit} - $p < 0$ (cf. Eq. (45), below)              \\
  \hline
   $\Gm \rarrow 1_{+}$    & ${\cal C} \rarrow - \infty$          & \texttt{isothermal limit} - $p > 0$ (cf. Eq. (45), below)              \\
  \hline
   $1 < \Gm < \gm$        & ${\cal C} < 0$                       & \texttt{quasi-adiabatic}                                               \\
  \hline
   $\Gm = \gm$            & ${\cal C} = 0$                       & \texttt{adiabatic}                                                     \\
  \hline
   $\gm < \Gm$            & $0 < {\cal C} < {\cal C}_V$          & \texttt{not physically-acceptable} (cf. Eq. (13), below)               \\
  \hline
   $\Gm \rarrow + \infty$ & ${\cal C} \rarrow {\cal C}_{V_-}$    & \texttt{not physically-acceptable} (cf. Eq. (13), below)               \\
  \hline
\end{tabular}
\vspace{.5cm}
\caption{Thermodynamic processes in an expanding Universe, along with the corresponding values of polytropic exponent, $\Gm$, and specific heat, ${\cal C}$.}
\end{table}

However, not all of these processes are physically-acceptable in an expanding Universe. For example, in a cosmological model with matter content in the form of perfect fluid, processes 9 and 10 of Table I are excluded, due to the second law of thermodynamics. In fact, according to this law, the entropy, ${\cal S}$, of a reversible process, obeying \be d{\cal S} = \frac{dQ}{T} \: , \ee is a never-diminishing function, i.e., \be \Dl {\cal S} = {\cal S}_{final} - {\cal S}_{initial} \geq 0 \: , \ee where the equality holds only for adiabatic processes. To determine the variation of entropy, along the polytropic transition of a perfect (cosmic) fluid from an initial state to a final one (the present epoch), we insert Eq. (1) into Eq. (8) and integrate, to obtain \be {\cal S}(t) - {\cal S}_0 = {\cal C} \ln \left [ \frac{T(t)}{T_0} \right ] \: , \ee where ${\cal S} (t)$ is the entropy of the fluid at some $t \leq t_0$ and ${\cal S}_0$ is the corresponding value at the present epoch $(t_0)$. In view of Eq. (4), Eq. (10) yields \be {\cal S} (t) = {\cal S}_0 + (\Gm - 1) {\cal C} \ln \left [ \frac{\rh (t)}{\rh_0} \right ] \ee and, by virtue of Eq. (6), it results in \be {\cal S} (t) = {\cal S}_0 + (\Gm - \gm) {\cal C}_V \ln \left [ \frac{\rh (t)}{\rh_0} \right ] \: . \ee In an expanding Universe, with matter in the form of (an, at least, calorically-) perfect fluid, ${\cal C}_V = ( \partial {\cal U} / \partial T)_V = constant > 0$ (see, e.g., Christians 2012), where ${\cal U}$ is the energy of this fluid's internal motions (per unit rest-mass). Furthermore, admitting that each volume element of the polytropic fluid is a closed thermodynamical system, total rest-mass (i.e., particles' number) consrvation implies that, for every $t \leq t_0$, $\rh (t) \geq \rh_0$. As a consequence, in view of Eq. (12), the second law of thermodynamics, given by Eq. (9), suggests that \be \Gm \leq \gm \ee and, hence, in a realistic polytropic cosmological model, the whole range of values $\Gm > \gm$ is excluded. According to the condition given by Eq. (13), in an expanding Univere filled with polytropic perfect fluid, there are only two physically-acceptable domains of values of the polytropic exponent, namely, \bea - \infty < \Gm < 1 & \Leftrightarrow & {\cal C} > 0 \Leftrightarrow dT > 0 \; \; \; and \\ 1 < \Gm \leq \gm & \Leftrightarrow & {\cal C} \leq 0 \Leftrightarrow dT < 0 \: , \eea separated by the isothermal $(\Gm = 1)$ barrier, for which $dT = 0$. Notice that, along with the expansion towards the present epoch, the temperature of a polytropic perfect fluid with $\Gm < 1$ increases $(dT > 0)$ and, therefore, so does its internal energy, as well.

On the other hand, the work (per unit mass) done by the pressure along the polytropic expansion of the cosmic fluid from an initial state (at which, $\rh = \rh (t)$) to a final one (the present epoch, at which $\rh = \rh_0 < \rh (t)$), is given by \be W_{\rh \rarrow \rh_0} = \int_{\rh}^{\rh_0} p dV = \int_{\rh}^{\rh_0} p d \left ( \frac{1}{\rh} \right ) \: , \ee which, by virtue of Eq. (3) (for $\Gm \neq 1$), yields \be W_{\rh \rarrow \rh_0} = \frac{1}{\Gm - 1} \frac{p_0}{\rh_0} \left [ \left ( \frac{\rh}{\rh_0} \right )^{\Gm - 1} - 1 \right ] \: . \ee As we shall clarify below (see e.g., Eq. (45)), the present-time value of the (conventional) pressure along a polytropic process in an expanding spatially-flat FRW model, is given by \be p_0 = \rh_0 c^2 (\Gm - 1) \frac{1 - \Om_M}{\Om_M} \: . \ee Hence, Eq. (17) results in \be W_{\rh \rarrow \rh_0} = \frac{1 - \Om_M}{\Om_M} c^2 \left [ \left ( \frac{\rh}{\rh_0} \right )^{\Gm - 1} - 1 \right ] \: . \ee In this case, for $\Gm < 1$ (i.e., $\Gm - 1 = - \vert 1 - \Gm \vert$), Eq. (19) is written in the form \be W_{\rh \rarrow \rh_0}^{\Gm < 1} = - \frac{1 - \Om_M}{\Om_M} c^2 \left [ 1 - \left ( \frac{\rh_0}{\rh} \right )^{\vert 1 - \Gm \vert} \right ] < 0 \: . \ee Since $\Om_M < 1$, Eq. (20) implies that, for $\Gm < 1$, the work done by the pressure (during the cosmic expansion of a polytropic perfect fluid) is negative, i.e., it is returned to the cosmic fluid itself.

Eventually, in a polytropic cosmological model, the first law of thermodynamics is given by \be d {\cal U} + p d \left ( \frac{1}{\rh} \right ) = {\cal C} d T \ee (see, e.g., Fock 1959, p. 83). In view of Eqs. (14) and (20), Eq. (21) suggests that, along a polytropic transition with $\Gm < 1$, the internal energy of the cosmic fluid increases $(d {\cal U} > 0)$, as a result of both $dT > 0$ and $dW = p d \left ( \frac{1}{\rh} \right ) < 0$. Hence, there (well) might be a point, at which, the thermodynamical-energy density (attributed to the internal motions of a polytropic cosmic fluid with $\Gm < 1$) dominates over the corresponding rest-mass quantity (i.e., $\rh {\cal U} > \rh c^2$), thus arising as a mighty contestant for the compensation of the extra (dark) energy amount, needed to flatten the Universe. In this context, a polytropic cosmological model could be a viable (and conventional) alternative to the DE concept, and, for this reason, in the Sections to follow, we shall scrutinize such a model. Nevertheless, it is not yet clear what kind of micro-physics would give the postulated polytropic behavior, and, therefore, our model is to be seen as an effective (phenomenological) approach of an elementary-physics scenario yet to be discovered.

\section{Polytropic processes in a cosmological (DM) fluid}

\subsection{The Universe matter-energy content}

The beginning of the $21^{st}$ century has been one of the most exciting epochs for cosmology as a science. According to the various CMB-oriented observational data, which became public at that time (see, e.g., de Bernardis et al. 2000; Jaffe et al. 2001; Padin et al. 2001; Stompor et al. 2001; Netterfield et al. 2002), the Universe can be adequately described by a spatially-flat RW model \be ds^2 = c^2 dt^2 - S^2 (t) \left ( dx^2 + dy^2 + dz^2 \right ) \: , \ee where $S (t)$ is the cosmic scale factor as a function of time. In such a model, the value of the Hubble parameter, $H(t)$, at the present epoch, is, by definition, given by \be H^2 (t_0) = \frac{8 \pi G}{3} \rh_c = H_0^2 \ee (see, e.g., Peacock 1999, p. 77). In the context of General Relativity (GR), the evolution of this model depends (also) on (the nature of) the source that drives the universal gravitational field, i.e., its matter-energy content. 

In specifying the Universe matter-energy content, we assume that (in principle) there is no DE at all. Instead, we admit that, the DM (along with the small, baryonic contamination) possesses fluid-like properties, in the sense that, the collisions of the WIMPs maintain a tight coupling between them and, hence, their (kinetic) energy can be re-distributed. In this case, the fundamental units of the Universe matter-energy content are the volume elements of the collisional-DM fluid (elements of fluid, each one consisting always of the same number of particles), which, as we (furthermore) assume, perform polytropic flows. 

In view of Bianchi identities, the motion of the volume elements in the interior of a gravitating continuous medium is governed by the equations \be T_{\; \; ; \nu}^{\mu \nu} = 0 \; , \ee where Greek indices refer to the four-dimensional spacetime (in connection, Latin indices refer to the three-dimensional spatial slices), the semicolon denotes covariant derivative, and $T^{\mu \nu}$ is the energy-momentum tensor of the Universe matter content, i.e., basically (but not solely), of the polytropic-DM fluid. Confining ourselves to the particular case of a perfect fluid, $T^{\mu \nu}$ takes on the standard form \be T^{\mu \nu} = (\varep + p)u^{\mu} u^{\nu} - p g^{\mu \nu} \: , \ee where $u^{\mu} = dx^{\mu}/ds$ is the four-velocity $\left ( u_{\mu}u^{\mu} = 1 \right )$ at the position of a fluid's volume element, $g^{\mu \nu}$ are the contravariant components of the Universe metric tensor, and $\varep$ is this fluid's total-energy density. Provided that internal structure is evident, in an (ideal) equilibrium state, $\varep$ can be decomposed according to \be \varep = w(\rh , T) + \rh {\cal U}(T) \ee (for a detailed analysis see, e.g., Fock 1959, pp. 81 - 83 and 91 - 94), where $\rh {\cal U}$ is the energy density associated with this fluid's thermodynamical content and $w$ denotes every other form of energy (density) involved. To determine ${\cal U}$ and $w$ (and, through them, $\varep$, as well), we address to thermodynamics and relativity, respectively.

\texttt{The first law of thermodynamics:} 

By virtue of Eqs. (3) and (4), the first law of thermodynamics (given by Eq. (21)) for polytropic flows yields \be {\cal U} = {\cal U}_0 \left ( \frac{\rh}{ \rh_0} \right )^{\Gm - 1} , \ee i.e., ${\cal U} \sim T$ (cf. Eq. (4)), where \be {\cal U}_0 = {\cal C} T_0 + \frac{1}{\Gm - 1} \frac{p_0}{\rh_0} \ee denotes the present-time value of the cosmic fluid's internal energy. Accordingly, the total-energy density of the Universe matter-energy content is written in the form \be \varep = w + \rh_0 {\cal U}_0 \left ( \frac{\rh}{\rh_0} \right )^{\Gm} . \ee  

\texttt{The continuity equation of GR:}

On the other hand, in terms of the metric tensor given by Eq. (22), the conservation law $T_{\; ; \nu}^{0 \nu} = 0$ reads \be \dot{\varep} + 3 \frac{\dot{S} }{S} (\varep + p) = 0 \: , \ee where the dot denotes differentiation with respect to cosmic time, $t$. Upon consideration of Eqs. (3) and (27) - (29), Eq. (30) results in \be \Gm {\cal U}_0 \left ( \dot{\rh} + 3 \frac{\dot{S}}{S} \rh \right ) + \dot{w} + 3 \frac{\dot{S}}{S} w - 3 (\Gm - 1) \rh_0 {\cal C} T_0 \frac{\dot{S}}{S} \left ( \frac{\rh}{\rh_0} \right )^{\Gm} = 0 \: . \ee At this point, we recall that, by definition, each volume element of the polytropic-DM fluid corresponds to a closed thermodynamical system, i.e., its particles' number is conserved. Consequently, \be \dot{\rh} + 3 \frac{\dot{S}}{S} \rh = 0 \: , \ee and, hence, \be \rh = \rh_0 \left ( \frac{S_0}{S} \right )^3 , \ee where $S_0$ is the value of $S (t)$ at the present epoch. In fact, Eq. (33) represents the conservation of the total rest-mass in a FRW cosmological model. In view of Eqs. (32) and (33), Eq. (31) is written in the form \be \dot{w} + 3 \frac{\dot{S}}{S} w - 3 (\Gm - 1) \rh_0 {\cal C} T_0 \frac{\dot{S}}{S} \left ( \frac{S_0}{S} \right )^{3 \Gm} = 0 \: , \ee i.e., as a linear differential equation of the first order to $w$, the solution of which reads \be w = \frac{A}{S^3} - \rh_0 {\cal C} T_0 \left ( \frac{S_0}{S} \right )^{3 \Gm} , \ee where $A$ is an integration constant. Taking into account Eq. (33), the dependence of the first term on the scale factor leads us to identify $A$ with the present-time value of the energy density corresponding to the total rest-mass, namely, \be A= \rh_0 c^2 S_0^3 \: . \ee Accordingly, Eq. (35) takes on its final form, as \be w = \rh_0 c^2 \left ( \frac{S_0}{S} \right )^3 - \rh_0 {\cal C} T_0 \left ( \frac{S_0}{S} \right )^{3 \Gm} , \ee that is, \be w = \rh c^2 - \rh {\cal C} T \: , \ee in which, the second term represents the heat per unit (of specific) volume, $Q / \left ( \frac{1}{\rh} \right )$, entering into the thermodynamical system. Eventually, by virtue of Eq. (37), Eq. (29) is written in the form \be \varep = \rh_0 c^2 \left ( \frac{S_0}{S} \right )^3 + \frac{p_0}{\Gm - 1} \left ( \frac{S_0}{S} \right )^{3 \Gm} = \rh c^2 + \frac{p}{\Gm -1} \: , \ee where we have used (also) Eq. (33). Upon consideration of Eq. (39), the dynamical evolution of the cosmological model given by Eq. (22), is governed by the Friedmann equation (with $\Lm = 0$) of the classical FRW Cosmology \be H^2 = \frac{8 \pi G}{3 c^2} \varep \: , \ee where \be H = \frac{\dot{S}}{S} \ee is the Hubble parameter as a function of the scale factor. However, there is an essential difference between this model and the rest of the classical FRW Cosmologies: In this case, the basic matter-constituents are the volume elements of a polytropic-DM fluid, i.e., they possess some sort of internal structure, and, therefore, thermodynamical content. As a consequence, the functional form of $\varep$ in Eq. (40) is no longer given by $\rh c^2$ alone, but by Eq. (39) (in connection, see Narlikar 1983, pp. 61, 62), and the Friedmann equation reads \be H^2 = \frac{8 \pi G}{3} \rh_0 \left ( \frac{S_0}{S} \right )^3 \left [ 1 + \frac{1}{\Gm - 1} \frac{p_0}{\rh_0 c^2} \left ( \frac{S_0}{S} \right )^{3 (\Gm - 1)} \right ] \: , \ee which, in view of Eq. (23), results in \be \left ( \frac{H}{H_0} \right )^2 = \Om_M \left ( \frac{S_0}{S} \right )^3 \left [ 1 + \frac{1}{\Gm - 1} \frac{p_0}{\rh_0 c^2} \left ( \frac{S_0}{S} \right )^{3 (\Gm - 1)} \right ] . \ee At the present epoch, when $S = S_0$ and $H = H_0$, Eq. (43) is reduced to \be \Om_M \left (1 + \frac{1}{\Gm - 1} \frac{p_0}{\rh_0 c^2} \right ) = 1 \: , \ee from which, the present-time value of the pressure arises, as \be p_0 = \rh_0 c^2 (\Gm - 1) \frac{1 - \Om_M}{\Om_M} \: , \ee and the Friedmann equation (43) is written in the form \be \left ( \frac{H}{H_0} \right )^2 = \left ( \frac{S_0}{S} \right )^3 \left [ \Om_M + (1 - \Om_M) \left ( \frac{S}{S_0} \right )^{3 (1 - \Gm)} \right ] \: . \ee For $\Gm < 1$, i.e., ${\cal C} > {\cal C}_V$ (see, e.g., Fig. 1), Eq. (45) suggests that, the (conventional) pressure given by Eq. (3) - refering to a gravitating perfect fluid which consists (mainly) of polytropic DM - is negative. In this case, the quantity $\varep + 3 p$ may also become negative (at some value of $S \leq S_0$), something that leads to $\ddot{S} > 0$ (see, e.g., Linder 2008; Caldwell \& Kamionkowski 2009). In other words, a cosmological model filled with a $(\Gm < 1)$ polytropic fluid, may accelerate its expansion (see Sect. 5.3, below). 

On the contrary, any cosmological model filled with matter in the form of a polytropic perfect fluid with $\Gm > 1$ (i.e., of positive pressure), is ever-decelerating. Indeed, for $\Gm \rarrow 1_{+}$, i.e., $\Gm = 1 + \eps$ with $\eps \rarrow 0$, Eq. (46) results in \be \left ( \frac{H}{H_0} \right )^2 = \left ( \frac{S_0}{S} \right )^3 \left [ 1 + 3 \eps (1 - \Om_M) \ln \left ( \frac{S_0}{S} \right ) + O (\eps^2) \right ] \: , \ee which, to linear terms in $\eps$, is of the same functional form as Eq. (24) of Kleidis \& Spyrou (2011), thus resulting in an ever-decelerating Universe. 

Furthermore, by virtue of Eq. (45), Eq. (39), for every value of $\Gm$, is written in the form \be \varep = \rh_0 c^2 \left [ \left ( \frac{S_0}{S} \right )^3 + \frac{1 - \Om_M}{\Om_M} \left ( \frac{S_0}{S} \right )^{3 \Gm} \right ] \: , \ee or, else, \be \varep = \rh_c c^2 \left [ \Om_M \left ( \frac{S_0}{S} \right )^3 + (1 - \Om_M) \left ( \frac{S_0}{S} \right )^{3 \Gm} \right ] > 0 \: . \ee Accordingly, in a cosmological model in which polytropic processes are dominant, the present-time (i.e., when $S = S_0$) value of the total-energy density, $\Om_0$, equals to unity, i.e., \be \Om_0 =  \frac{\varep_0}{\varep_c} = \frac{\rh_c c^2}{\rh_c c^2} \left [ \Om_M + (1 - \Om_M) \right ] = 1 \: . \ee In view of Eqs. (49) and (50), the extra (dark) energy, needed to compromise spatial flatness of the cosmological model given by Eq. (22), can be provided by the energy of the internal motions of a collisional-DM fluid, the volume elements of which perform polytropic flows. Hence, a Universe with matter content in the form of a polytropic-DM fluid with $\Gm < 1$, might be a relatively inexpensive solution to the whole DE concept, in the sense that, it can address both the extra energy needed for spatial flatness and the subsequent accelerated expansion, in one single model. For this reason, in what follows, we shall focus on a cosmological model filled with a polytropic (DM) perfect fluid with $\Gm < 1$.

In such a model, by virtue of Eq. (48), we can identify the rest-mass energy density, $\varep_{mat} = \rh c^2$, and the extra (dark) energy density, $\varep_{int} = \varep - \varep_{mat}$, of the Universe total matter-energy content and express these quantities as functions of the cosmological redshift parameter, \be z + 1 = \frac{S_0}{S} \: , \ee yielding \be \frac{\varep_{int}}{\varep_{mat}} = \frac{1 - \Om_M}{\Om_M} \frac{1}{(1+z)^{3(1 - \Gm)}} \: . \ee At the present epoch $(z = 0)$, and for every value of $\Gm$, Eq. (52) results in \be \left . \frac{\varep_{int}}{\varep_{mat}} \right \vert_0 = \frac{1 - \Om_M}{\Om_M} \: , \ee as it should (in a spatially-flat Universe). 

Notice that, for $\Om_M = 0.274$ (Komatsu et al. 2011), Eq. (45) suggests that, today, $p_0 = - 2.650 (1 - \Gm) \rh_0 c^2$. This result could mislead even the careful reader, to assume that the polytropic cosmological model is nothing but a phantom Universe (where $p_0 < - \varep_0$). However, we need to stress that, by virtue of Eq. (48), in our (polytropic) cosmological model, the (total) energy density at the present epoch is not given by $\rh_0 c^2$, but by $\varep_0 = \Om_M^{-1} \rh_0 c^2$. Accordingly, Eq. (45) results in \be p_0 = - (1 - \Gm) (1 - \Om_M) \varep_0 \: . \ee In this case, as long as \be (1 - \Gm) (1 - \Om_M) < 1 \Leftrightarrow \Gm > - \frac{\Om_M}{1 - \Om_M} \cong - 0.377 \: , \ee we obtain \be p_0 > - \varep_0 \: , \ee and, therefore, the polytropic-DM model no longer belongs to the realm of phantom Cosmology. Indeed, as we shall demonstrate later on, Eq. (55) is valid in a cosmological model, in which the accelerated expansion takes place at a lower rate than de Sitter expansion (cf. also Eq. (75)).

\subsection{The Universe scale factor and the cosmic time}

In a cosmological model filled with a polytropic perfect fluid, Eq. (46) yields \bea && \left [ \frac{d}{d t} \left ( \frac{S}{S_0} \right )^{3/2} \right ]^2 = \nn \\ && \frac{1}{t_{EdS}^2} \left \lbrace \Om_M + (1 - \Om_M) \left [ \left ( \frac{S}{S_0} \right )^{3/2} \right ]^{2 (1 - \Gm)} \right \rbrace \: , \eea where \be t_{EdS} = \frac{2}{3 H_0} \ee is the age of the Universe of the Einstein-de Sitter (EdS) model (dust Universe). In order to solve Eq. (57), we set \be 0 \leq \ch = \left ( \frac{S}{S_0} \right )^{3/2} \leq 1 \: . \ee In accordance, Eq. (57) results in \be \int_0^{\ch} \frac{d \ch}{\sqrt{\Om_M + (1 - \Om_M) \ch^{2 (1 - \Gm)}}} = \frac{t}{t_{EdS}} \: . \ee Eq. (60) can be solved explicitly in terms of hypergeometric functions, $_2F_1 (a \: , \: b \: ; \: c \: ; \: x)$, of a complex variable, $x$ (see, e.g., Gradshteyn \& Ryzhik 2007 (7th Ed.), pp. 1005 - 1008), as follows \bea && \left ( \frac{S}{S_0} \right )^{\frac{3}{2}} \times \nn \\ && _2F_1 \left ( \frac{1}{2 (1 - \Gm)} \: , \: \frac{1}{2} \: ; \: \frac{3 - 2 \Gm}{2 (1 - \Gm)} \: ; - \left ( \frac{1 - \Om_M}{\Om_M} \right ) \left [ \frac{S}{S_0} \right ]^{3 (1 - \Gm)} \right ) \nn \\ && = \sqrt{\Om_M} \left ( \frac{t}{t_{EdS}} \right ) , \eea which, for $\Om_M = 1$, yields $S = S_0 \left ( \frac{t}{t_{EdS}} \right )^{2/3}$, i.e., the EdS model, as it should. Since $a + b = \frac{1}{2 (1 - \Gm)} + \frac{1}{2} < \frac{3 - 2 \Gm}{2 (1 - \Gm)} = c$, the hypergeometric series involved in Eq. (61), converges absolutely within the unit circle $\left \vert \frac{S}{S_0} \right \vert \leq 1$, for every value of $\Gm < 1$ (see, e.g., Abramowitz \& Stegun 1970, p. 556). It is worthnoting that, in the isobaric $\Gm = 0$ case, Eq. (61) is reduced to \bea &&\left ( \frac{S}{S_0} \right )^{\frac{3}{2}} \: _2F_1 \left ( \frac{1}{2} \: , \: \frac{1}{2} \: ; \: \frac{3}{2} \: ; \: - \left ( \frac{1 - \Om_M}{\Om_M} \right ) \left [ \frac{S}{S_0} \right ]^3  \right ) \nn \\&&= \sqrt{\Om_M} \left ( \frac{t}{t_{EdS}} \right ) \: , \eea which, upon consideration of the identity \be _2F_1 \left ( \frac{1}{2} \: , \: \frac{1}{2} \: ; \: \frac{3}{2} \: ; \: - x^2  \right ) = \frac{1}{x} \sinh^{-1} (x) \ee (cf. Abramowitz \& Stegun 1970, Eq. (15.1.7), p. 556; Gradshteyn \& Ryzhik 2007 (7th Ed.), Eq. 9.121.28, p. 1007), results in \be S(t) = S_0 \left ( \frac{\Om_M}{1 - \Om_M} \right )^{1/3} \sinh^{2/3} \left ( \sqrt{1 - \Om_M} \frac{t}{t_{EdS}} \right ) \: , \ee i.e., in a functional form similar to the corresponding $\Lm$CDM result. Moreover, from Eq. (61) we may determine the age of the Universe, $t_0$ (i.e., the time at which $S = S_0$), in a polytropic-DM model. In units of $t_{EdS}$, it is given by \bea && \frac{t_0}{t_{EdS}} = \frac{1}{\sqrt {\Om_M}} \times \nn \\ && _2F_1 \left ( \frac{1}{2 (1 - \Gm)} \: , \: \frac{1}{2} \: ; \: 1 + \frac{1}{2 (1 - \Gm)} \: ; \: - \frac{1 - \Om_M}{\Om_M} \right ) \: , \eea the behavior of which, as a function of the polytropic exponent, $\Gm < 1$, is presented in Fig. 2. For $p = constant = p_0$ (i.e., $\Gm = 0$), Eq. (65) yields \be t_0 = t_{EdS} \frac{1}{\sqrt{1 - \Om_M}} \sinh^{-1} \sqrt{\frac{1 - \Om_M}{\Om_M}} \: , \ee which, for $\Om_M = 0.274$ (Komatsu et al. 2011), results in $t_0 = 1.483 \; t_{EdS} = 13.773 \; Gys$, i.e., the age of the Universe of the $\Lm$CDM model. 

\begin{figure}[ht!]
\centerline{\mbox {\epsfxsize=9.cm \epsfysize=7.cm
\epsfbox{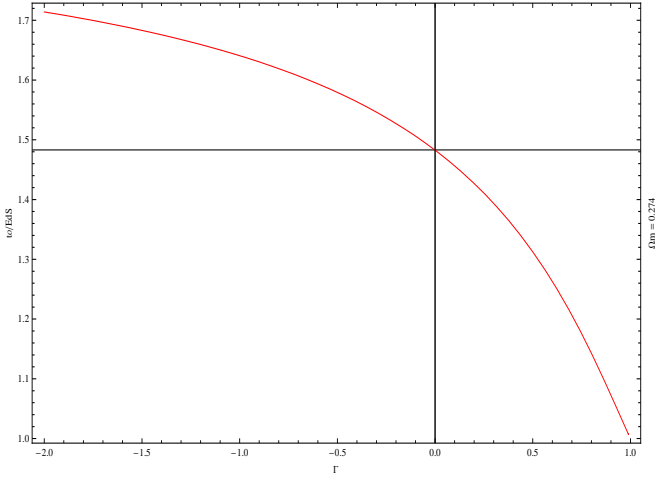}}} \caption{Age of a polytropic-DM model, $t_0$, in units of $t_{EdS}$, as a function of the polytropic exponent $\Gm < 1$ (red solid line). Notice that, for every $\Gm < 1$, $t_0 > t_{EdS}$, and, in fact, $t_0$ approaches $t_{EdS}$ only as $\Gm \rarrow 1$. The horizontal solid line denotes the age of the Universe, $t_0 = 1.483 \; t_{EdS}$, in the $\Lm$CDM-limit of the polytropic-DM model (i.e., for $\Gm = 0$).}
\end{figure}

Eventually, combining Eqs. (61) and (65), we obtain the equation that governs the time evolution of a polytropic-DM model. It is written in the form \bea && \left ( \frac{S}{S_0} \right )^{3/2} \times \nn \\ && \frac{_2F_1 \left ( \frac{1}{2 (1 - \Gm)} \: , \: \frac{1}{2} \: ; \: \frac{3 - 2 \Gm}{2 (1 - \Gm)} \: ; \: - \left ( \frac{1 - \Om_M}{\Om_M} \right ) \left [ \frac{S}{S_0} \right ]^{3 (1 - \Gm)} \right )}{_2F_1 \left ( \frac{1}{2 (1 - \Gm)} \: , \: \frac{1}{2} \: ; \: \frac{3 - 2 \Gm}{2 (1 - \Gm)} \: ; \: - \frac{1 - \Om_M}{\Om_M} \right )} \nn \\ && = \frac{t}{t_0} \: . \eea In this case, the time behavior of the cosmic scale factor, for several values of the polytropic exponent $\Gm < 1$, is presented in Fig. 3. The profiles of the corresponding curves suggest that, there is always a value of $t < t_0$, above which, the function $S(t)$ becomes concave (i.e., $\ddot{S} > 0$); in other words, a cosmological model filled with a polytropic (DM) perfect fluid, for $\Gm < 1$, accelerates its expansion. This can be (most appropriately) confirmed, if someone calculates the corresponding deceleration parameter, $q$.

\begin{figure}[ht!]
\centerline{\mbox {\epsfxsize=9.cm \epsfysize=7.cm
\epsfbox{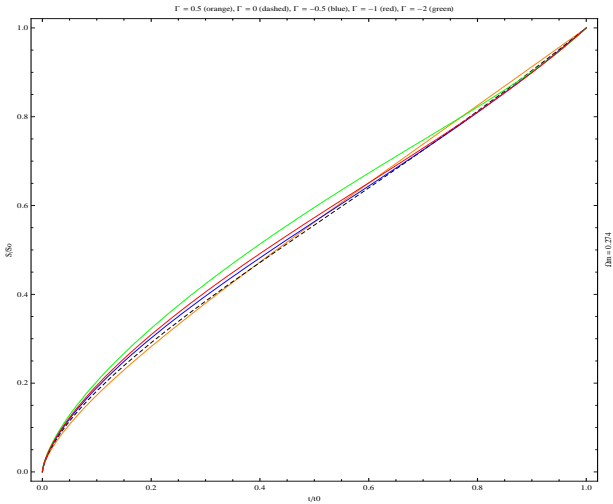}}} \caption{Scale factor, $S$, of a cosmological model driven by a polytropic-DM fluid with $\Om_M = 0.274$ (in units of its present-time value, $S_0$), as a function of the cosmic time $t$ (in units of $t_0$), for $\Gm = 0.5$ (orange), $\Gm = 0$ (dashed), $\Gm = -0.5$ (blue), $\Gm = - 1$ (red) and $\Gm = - 2$ (green). Notice that, there is always a value of $t < t_0$, above which $\ddot{S} > 0$, i.e., the Universe accelerates its expansion.}
\end{figure}

\subsection{Deceleration parameter and the subsequent accelerated expansion}

In a cosmological model, in which the polytropic flow (with $\Gm < 1$) of the cosmic fluid's volume elements is the dominant type of motion, the Hubble parameter (46), in terms of the cosmological redshift parameter, $z$, is written in the form \be H = H_0 ( 1 + z )^{\frac{3}{2}} \left [ \Om_M + \frac{1 - \Om_M}{( 1 + z )^{3 (1 - \Gm)}} \right ]^{1/2} . \ee Accordingly, the corresponding deceleration parameter, in terms of $H$ and $z$, is given by \be q (z) = \frac{dH / dz}{H(z)} (1+z) - 1 \: , \ee which, upon consideration of Eq. (68), yields \be q (z) = \frac{1}{2} \left [ 1 - \frac{3 (1 - \Gm) (1 - \Om_M)}{\Om_M (1 + z)^{3 (1 - \Gm)} + (1 - \Om_M)} \right ] \: . \ee For $z \gg 1$ (i.e., at the distant past), $q \rarrow \frac{1}{2}$ and the Universe behaves as a matter (dust) dominated (in other words, decelerating) FRW model. On the other hand, for $z = 0$ (i.e., at the present epoch), we have \be q_0 = \frac{1}{2} \left [ 1 - 3 (1 - \Gm) (1 - \Om_M) \right ] \: , \ee which, depending on $\Gm$, can be either positive or negative or zero. In fact, the minus sign on the rhs of Eq. (70) suggests that, there is a transition value of $z$, namely, $z_{tr}$, below which, $q(z)$ does become negative, i.e., the Universe accelerates its expansion. In a polytropic-DM model, $z_{tr}$ is a function of the polytropic exponent, given by \be z_{tr} = \left [ (2 - 3 \Gm) \frac{1 - \Om_M}{\Om_M} \right ]^{\frac{1}{3 (1 - \Gm)}} - 1 \: . \ee Notice that, in view of Eq. (6), in a polytropic fluid with ${\cal C} > {\cal C}_V$ (i.e., $\Gm < 1$), the condition \be \Gm = \frac{\gm - \frac{{\cal C}}{{\cal C}_V}}{1 - \frac{{\cal C}}{{\cal C}_V}} < \frac{2}{3} \ee leads to $\gm - \frac{2}{3} > \frac{1}{3} \left ( \frac{{\cal C}}{{\cal C}_V} \right ) > \frac{1}{3} \Rarrow \gm > 1$ which is valid, anyway. Hence, in what follows we consider $\Gm < \frac{2}{3}$ (in connection, see also Freese \& Lewis 2002; Gondolo \& Freese 2003). In this context, the condition $z_{tr} \geq 0$ (equivalently, $q_0 \leq 0$) imposes a further constraint on $\Gm$ itself, namely, \be \Gm \leq \frac{1}{3} \left [ 2 - \frac{\Om_M}{1 - \Om_M} \right ] \: , \ee which, for $\Om_M = 0.274$ (Komatsu et al. 2011), yields an upper limit of $\Gm$, namely, $\Gm \leq 0.541$. On the other hand, admitting that, the accelerating Universe will approach a de Sitter phase only in the distant future, the condition $q_0 > -1$ (equivalently, $p_0 > - \varep_0$) should (also) be imposed, leading to \be \Gm > - \frac{\Om_M}{1 - \Om_M} \cong - 0.377 \: , \ee which may serve as a lower bound of $\Gm$ (cf. also Eq. (55)). Finally, for $\Gm = 0$, Eq. (72) yields $z_{tr} = 0.744$, which lies (well) within the current range of the corresponding $\Lm$CDM result, namely, $z_{tr} = 0.752 \pm 0.041$ (see e.g., Suzuki et al. 2012). The behavior of $z_{tr}$, as a function of $\Gm \leq 0.541$, is presented in Fig. 4.

\begin{figure}[ht!]
\centerline{\mbox {\epsfxsize=9.cm \epsfysize=7.cm
\epsfbox{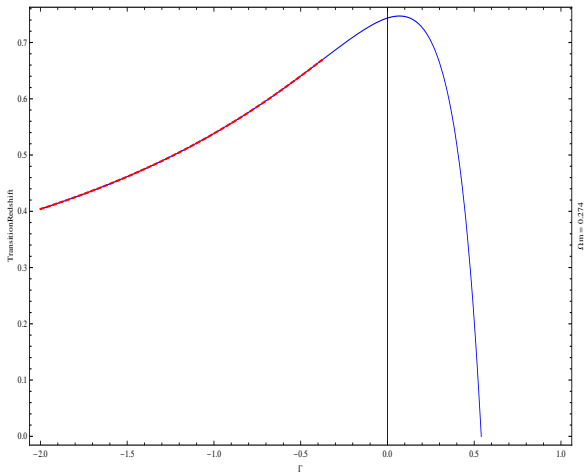}}} \caption{Transition redshift, $z_{tr}$, in a polytropic-DM model as a function of the polytropic exponent, $\Gm$ (blue solid curve). Notice that, the constraint $z_{tr} \geq 0$ yields $\Gm \leq 0.541$, while, the condition $q > -1$ results in $\Gm > -0.377$, below which the Universe enters in the phantom realm (red dashed curve).}
\end{figure}

In view of the aforementioned (theoretical) resutls, polytropic acceleration is (certainly) not a coincidence. In other words, a cosmological model filled with a polytropic (DM) perfect fluid of $- 0.377 < \Gm \leq 0.541$, (most naturally) accelerates its expansion at cosmological redshifts lower than a transition value, given by Eq. (72), without the need for either any exotic DE or the cosmological constant. 

The question that arises now, is, whether (or not) these results can be confirmed (also) by observational data of cosmological significance, especially, those that led to the assumption of the accelerating expansion in the first place. In fact, as we demonstrate in the next Section, a Universe with matter content in the form of a polytropic-DM fluid, can reproduce (to high accuracy) the observational distribution of the SNe Ia distant indicators.

\section{Accomodating the SN Ia observational data}

Nowadays, the most direct and reliable method for determining, observationally, the (relatively) recent history of the Universe expansion, is to measure the redshift and the apparent luminosity (equivalently, the apparent magnitude, $m$) of cosmologically-distant indicators (standard candles), whose absolute luminosity (equivalently, the absolute magnitude, $M$) is assumed to be known.

SN Ia events constitute one of the most suitable cosmological standard candles. Today, more than 600 SN Ia events have been identified spectroscopically (see, e.g., Suzuki et al. 2012) by a number of scientific groups (see, e.g., Hamuy et al. 1996; Garnavich et al. 1998; Perlmutter et al. 1998, 1999$a$; Riess et al. 1998, 2001, 2004, 2007; Schmidt et al. 1998; Knop et al. 2003; Tonry et al. 2003; Barris et al. 2004; Krisciunas et al. 2005; Astier et al. 2006; Jha et al. 2006; Miknaitis et al. 2007; Wood-Vasey et al. 2007; Amanullah et al. 2008, 2010; Holtzman et al. 2008; Kowalski et al. 2008; Hicken et al. 2009$a$, 2009$b$; Kessler et al. 2009; Contreras et al. 2010; Guy et al. 2010; Suzuki et al. 2012). In each and every one of these surveys, the SN Ia events (at peak luminosity) appear to be dimmer (i.e., they seem to lie farther away) than what would have been expected in the context of a dust (i.e., pressureless) Universe. 

At this point, we recall that, when spatial flatness was established, the common perception about the cosmos was that the DM constituents are collisionless (hence the Universe matter content was most favorably interpreted as dust), something that (necessarily) led to the assumption of an extra (dark) energy component. However, in a polytropic-DM model, such an assumption would have not been necessary, since, the appropriate candidate to provide the extra energy (needed to flatten the Universe) would have already been included in the model (the energy of the internal motions). In this context, we cannot help but wondering, whether (also) the observed distribution of the SNe Ia standard candles can be appropriately accomodated in a polytropic-DM model, or not. 

Many samples of SN Ia data have already appeared in the literature, in order to scrutinize the viability of the various DE scenarios (see, e.g., Davis et al 2007; Cuhna 2009). In this context, today, there is the most extended SN Ia dataset, consisting of 580 events, which is known as the Union 2.1 Compilation (Suzuki et al. 2012). This sample\footnote{Available at \texttt{http://www.supernova.lbl.gov/Union}} is an augmented version of the Union 2 SN Ia Compilation (Amanullah et al. 2010), based on the systematic methodology developed by Kowalski et al. (2008). In addition to the Union 2 SN dataset, the Union 2.1 Compilation includes 102 low-redshift SNe Ia from the CfA3 survey (Hicken et al. 2009$a$), 129 intermediate-redshift SNe Ia (Holtzman et al. 2008), 5 intermediate-redshift events discovered from La Palma (Amanullah et al. 2008), and 6 new high-redshift SNe Ia data (Suzuki et al. 2012). 

In order to estimate the contribution of the polytropic approach to the DE concept, we shall overplot the corresponding theoretically-derived distance modulus, $\mu (z) = m - M$, on the Hubble ($\mu$ versus $z$) diagram of the Union 2.1 SN Compilation. The K-corrected distance modulus of a light-emitting source is given by \be \mu (z) = 5 \log \left ( \frac{d_L}{Mpc} \right ) + 25 \ee (see, e.g., Narlikar 1983, Eqs. (13.10) and (13.12), p. 359), where $d_L$ is the luminosity distance of the source, measured in megaparsecs $(Mpc)$. In a spatially-flat model, $d_L$ can be expressed as a function of the cosmological redshift and the Hubble parameter, as follows \be d_L (z) = c (1+z) \int_0^z \frac{d z^{\prime}}{H(z^{\prime})} \ee (see, e.g., Peacock 1999, p. 76). Accordingly, inserting Eq. (68) into Eq. (77), we obtain \bea && d_L (z) = \frac{c}{H_0} (1+z) \times \nn \\ && \int_0^z \frac{d z^{\prime}}{ \left ( 1 + z^{\prime} \right )^{\frac{3 \Gm}{2}} \left [ (1 - \Om_M) + \Om_M \left ( 1 + z^{\prime} \right )^{3 (1 - \Gm)} \right ]^{1/2}} \: . \eea Once again, the integral on the rhs of Eq. (78) can be solved explicitly in terms of hypergeometric functions (see e.g., Gradshteyn \& Ryzhik 2007 (7th Ed.), pp. 1005 - 1008), as follows \bea && d_L (z) = \frac{2 c}{H_0} \frac{1}{\sqrt {1 - \Om_M}} \frac{1 + z}{2 - 3 \Gm} \left [ (1 + z)^{\frac{2 - 3 \Gm}{2}} \times \right . \nn \\ && \left . _2F_1 \left ( \frac{2 - 3 \Gm}{6 ( 1 - \Gm)}  \: , \: \frac{1}{2} \: ; \: \frac{8 - 9 \Gm}{6 ( 1 - \Gm)} \: ; \: - \left [ \frac{\Om_M}{1 - \Om_M} \right ] (1 + z)^{3 (1 - \Gm)} \right )  - \right . \nn \\ && \left . _2F_1 \left ( \frac{2 - 3 \Gm}{6 (1 - \Gm)}  \: , \: \frac{1}{2} \: ; \: \frac{8 - 9 \Gm}{6 ( 1 - \Gm)} \: ; \: - \left [ \frac{\Om_M}{1 - \Om_M} \right ] \right ) \right ] \: . \eea 

\begin{figure}[ht!]
\centerline{\mbox {\epsfxsize=9.cm \epsfysize=7.cm
\epsfbox{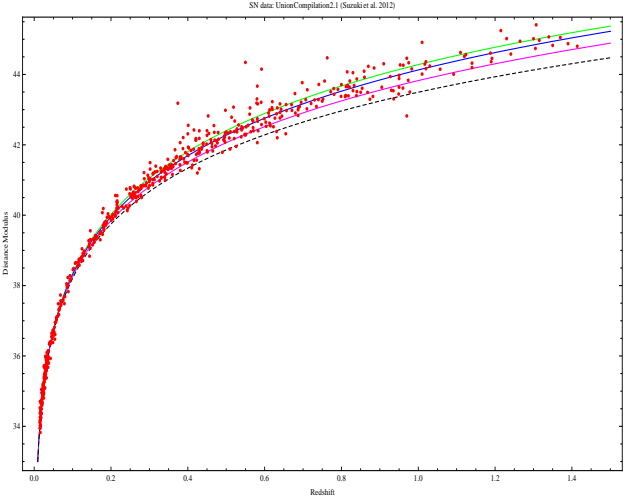}}} \caption{Hubble diagram of the Union 2.1 SN Compilation (red dots). Overplotted are the theoretical curves, representing the distance modulus as a function of the cosmological redshift in the polytropic-DM model, for $\Om_M = 0.274$ and several (acceptable) values of the polytropic exponent, namely, $\Gm = 0.541$ (magenta), $\Gm = 0$ (blue), and $\Gm = -0.377$ (green), as compared to the corresponding quantity in the collisionless-DM (dust) case (dashed curve).}
\end{figure}

\noindent Next, we overplot Eq. (76), with $d_L (z)$ being given by Eq. (79), on the $\mu$ versus $z$ diagram of 580 SNe Ia, of the extended Union 2.1 Compilation (Suzuki et al. 2012). To do so, we follow Komatsu et al. (2011), admitting that, today, $\Om_M = 0.274$ and $H_0 = 70.2$ $Km/sec/Mpc$; hence, $2c/H_0 = 8,547$ $Mpc$. The outcome is presented in Fig. 5, for several values of the polytropic exponent, $-0.377 < \Gm \leq 0.541$. As expected, the theoretically-derived curves representing the distance modulus, $\mu$, as a function of $z$ in the polytropic-DM model, fit the entire SN Ia distribution much more accurately than the collisionless-DM (EdS) formula (dashed curve), given by \be d_L^{EdS} (z) = \frac{2c}{H_0} \: \left ( 1 + z \right )^{1/2} \: \left [ \left ( 1 + z \right )^{1/2} - 1 \right ] \ee (see, e.g., Carroll et al. 1992). In this case, from Fig. 6, we see that, the best fit of the theoretically-derived result, given by Eqs. (76) and (79), to the observationally-determined Hubble diagram of the Union 2.1 SN dataset is achieved for $-0.089 < \Gm \leq 0$ (in connection, see next Section). 

\begin{figure}[ht!]
\centerline{\mbox {\epsfxsize=9.cm \epsfysize=7.cm
\epsfbox{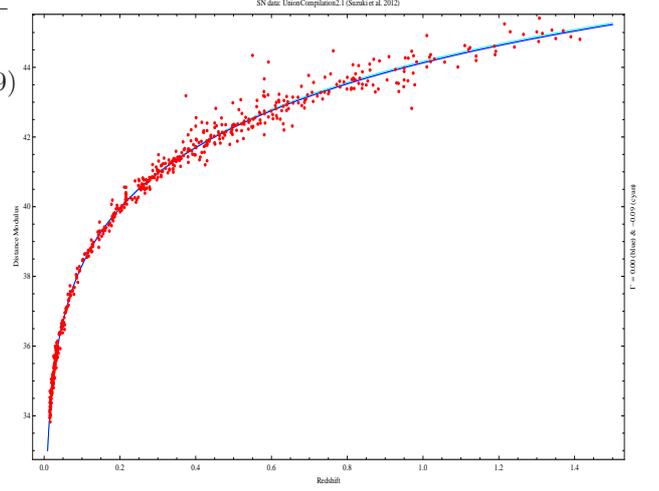}}} \caption{Overplotted to the observationally-determined Hubble diagram of the Union 2.1 Compilation, are the best-fit curves (too close to be distinguished) representing the function $\mu (z)$ in the polytropic-DM model, when $-0.089 < \Gm \leq 0$.}
\end{figure}

\section{The physics of transition}

\subsection{The velocity of sound}

To determine the velocity of sound in relativistic hydrodynamics, one simply begins with the conservation law, given by Eq. (24), and monitors its response to infinitesimal compressions (or expansions) of the fluid. To linear perturbation-terms, the resulting wave equation \be \frac{\partial^2 \dl}{\partial t^2} - c^2 \left ( \frac{\partial p}{\partial \varep} \right )_{\cal S} \nabla^2 \dl = 0 \: , \ee where $\dl = \frac{\dl n}{n}$ is the particles-number density-contrast, defines the isentropic velocity of sound as \be c_s^2 = c^2 \left ( \frac{\partial p}{\partial \varep} \right )_{\cal S} \ee (see, e.g., Weinberg 1972, p. 52). In view of Eq. (82), barotropic flow in a polytropic-DM perfect fluid, defined by Eqs. (3), (33) and (39), suggests that, in the cosmological model under consideration, the velocity of sound may no longer be constant, but a function of the cosmological redshift, parametrized by $\Gm$. Accordingly, we distinguish two cases.

\texttt{(i)} $\Gm = 0$: In this case, $p = p_0 = constant$, and, therefore, \be c_s^2 (\Gm = 0) = 0 \: . \ee In other words, in an isobaric cosmological model, no acoustic waves ever propagate; the Universe remains "silent".

\texttt{(ii)} $\Gm \neq 0$: In this case, by virtue of Eqs. (3), (33) and (39), the total-energy density of the Universe matter-energy content is written in the form \be \varep = \underbrace{\rh c^2}_{\varep_{mat}} + \underbrace{\frac{p}{\Gm - 1}}_{\varep_{int}} = \rh_0 c^2 \left (\frac{p}{p_0} \right )^{1/\Gm} + \frac{p}{\Gm - 1} \: . \ee Partial differentiation of Eq. (84) with respect to $\varep$, yields \be \left ( \frac{\partial p}{\partial \varep} \right )_{\cal S} = \frac{\Gm \left ( \frac{p}{\rh c^2} \right )}{1 + \frac{\Gm}{\Gm - 1} \left ( \frac{p}{\rh c^2} \right )} = \left ( \frac{c_s}{c} \right )^2 \: . \ee Accordingly, the velocity of sound as a function of the cosmological redshift, is given by \be \left ( \frac{c_s}{c} \right )^2 =  - \frac{\Gm (1 - \Gm) \frac{1 - \Om_M}{\Om_M}}{(1 + z)^{3(1 - \Gm)} + \Gm \frac{1 - \Om_M}{\Om_M}} \: . \ee In view of Eq. (55), the denominator on the rhs of Eq. (86) is positive for every $z \geq 0$, and, therefore, the condition of a positive velocity-of-sound square yields a major constraint on the upper bound of $\Gm$, namely, \be \left ( \frac{c_s}{c} \right )^2 > 0 \Leftrightarrow \Gm < 0 \: . \ee Hence, in what follows, we consider $\pm \Gm = \mp \vert \Gm \vert$. Now, Eq. (86) is written in the form \be \left ( \frac{c_s}{c} \right )^2 = \left ( 1 + \vert \Gm \vert \right ) \frac{\vert \Gm \vert \frac{1 - \Om_M}{\Om_M}}{(1 + z)^{3(1 + \vert \Gm \vert)} - \vert \Gm \vert \frac{1 - \Om_M}{\Om_M}} \: . \ee 

With respect to $z$, there are two values of $\left ( \frac{c_s}{c} \right )^2$ of particular interest, namely, \texttt{(a)} at transition $(z = z_{tr})$, where \be \left ( \frac{c_s}{c} \right )_{tr}^2 = \frac{\vert \Gm \vert}{2} \Rightarrow \vert \Gm \vert = 2 \left ( \frac{c_s}{c} \right )_{tr}^2 \: , \ee attributing to the polytropic exponent an unexpected physical interpetation, and \texttt{(b)} at the present epoch $(z = 0)$, when $\left ( \frac{c_s}{c} \right )^2$ attains its maximum value, namely, \be \left ( \frac{c_s}{c} \right )_0^2 = \left ( 1 + \vert \Gm \vert \right ) \frac{\vert \Gm \vert \frac{1 - \Om_M}{\Om_M}}{1 - \vert \Gm \vert \frac{1 - \Om_M}{\Om_M}} \: . \ee At this point, we recall that, for relativistic particles (i.e., HDM), the velocity of sound reads $\left ( \frac{c_s}{c} \right )^2 = \frac{1}{3}$ (see, e.g., Weinberg 1972, p. 51; Landau \& Lifshitz 1987, p. 509). Accordingly, the cosmological requirement for CDM at the present epoch, is translated to \be \left ( \frac{c_s}{c} \right )_0^2 < \frac{1}{3} \: , \ee which, in the polytropic-DM model under consideration, results in \be \vert \Gm \vert^2 \frac{1 - \Om_M}{\Om_M} + \frac{4}{3} \vert \Gm \vert \frac{1 - \Om_M}{\Om_M} - \frac{1}{3} < 0 \: , \ee yielding \be \vert \Gm \vert < \frac{2}{3} \left [ \sqrt{1 + \frac{3}{4} \frac{\Om_M}{1 - \Om_M}} - 1 \right ] \: \overbrace{=}^{\Om_M = 0.274} 0.089 \: . \ee Notice that, the (reasonable) physical requirements given by Eqs. (87) and (91), together with Eq. (83), have led to an even narrower range of values of the only free parameter of this model, namely, \be - 0.089 < \Gm \leq 0 \: , \ee i.e., in a realistic polytropic-DM cosmological model, the polytropic exponent - if not zero (i.e., a $\Lm$CDM-equivalent model) - is definitely negative and very close to zero (in connection, see Fig. 6). The result given by Eq. (94) lies well-within the corresponding range obtained for a generalized Chaplygin gas $(p \sim - \rh^{- \al})$, from the analysis of X-ray and SN Ia measurements, in connection to data from Fanaroff-Riley type IIb radio-galaxies, namely, $\al = -0.09_{-0.33}^{+0.54}$ (Zhu 2004). 

In view of Eq. (94), Eq. (89) yields \be 0 \leq \left ( \frac{c_s}{c} \right )_{tr}^2 < 0.044 \: , \ee while, by virtue of Eq. (71), the present-time value of the deceleration parameter keeps to \be -0.686 < q_0 \leq -0.589 \: . \ee Eq. (96) has a well-shaped cross-section with the lower part of the observationally-determined range of values of $q_0$ (based on the SALT2 fitting to the SNe+BAO/CMB data), namely, $q_0 = -0.53_{-0.13}^{+0.17}$ (Giostri et al. 2012). 

Furthermore, the combination of Eqs. (6) and (94) results in a most definite set of (allowed) values of the heat capacity, ${\cal C}$ (in units of ${\cal C}_V$), in a polytropic (DM) Universe, namely, those in the range \be 0.082 + 0.918 \gm < \frac{\cal C}{{\cal C}_V} \leq \gm \: , \ee which, for $1 < \gm \leq 3$ (see, e.g., Guidry 1998, p. 131), is depicted in Fig. 7 (blue-filling region).

\begin{figure}[ht!]
\centerline{\mbox {\epsfxsize=9.cm \epsfysize=7.cm
\epsfbox{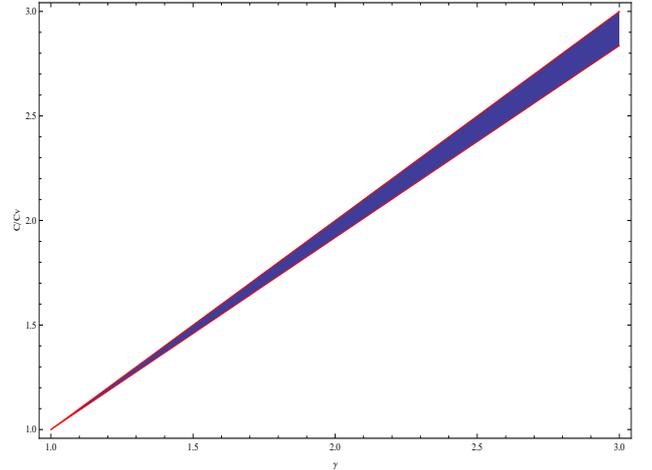}}} \caption{Allowed set of values of the heat capacity ${\cal C}$ (in units of ${\cal C}_V$) in a polytropic-DM model with $-0.089 < \Gm \leq 0$ (blue-filling region), as a function of the adiabatic index $1 < \gm \leq 3$.}
\end{figure}

\subsection{Observables at transition}

As we saw in Sect. 2, for $\Gm < 1$, the work done by the pressure along a polytropic process in an expanding Universe, is given back to the cosmic fluid itself. As a consequence, both the internal energy, ${\cal U}$, and the temperature, $T$, of the polytropic-DM fluid increase towards "now". It is straightforward to prove that the change of ${\cal U}$ and $T$ from some $z$ to $z = 0$ is given by \bea \left. \frac{\Dl {\cal U}}{\cal U} \right \vert_{z \rarrow 0} & = & \frac{{\cal U}_0 - {\cal U}(z)}{{\cal U}_0} = 1 - \frac{1}{(1+z)^{3(1+ \vert \Gm \vert)}} \nn \\ & = & \frac{T_0 - T(z)}{T_0} = \left. \frac{\Dl T}{T} \right \vert_{z \rarrow 0} \: , \eea which, at $z = z_{tr}$, results in \be \frac{T_{tr}}{T_0} = \left ( \frac{\Om_M}{1 - \Om_M} \right ) \frac{1}{2 + 3 \vert \Gm \vert} = \frac{0.377}{2 + 3 \vert \Gm \vert} \: . \ee From Eq. (99), we find that, for $-0.089 < \Gm \leq 0$, the temperature of DM at transition, in units of its present-time value, lies in the range \be 0.165 < \frac{T_{tr}}{T_0} \leq 0.189 \: , \ee a prediction to be verified by observation. Of course, the exact value of $T_0$ depends on the exact nature of the DM constituents. Notice, however, that, $T_0$ cannot exceed the present-time cosmic neutrino background temperature, $T_{\nu} = 1.95 \; K$, in order for the Big Bang nucleosynthesis to remain unaffected (see, e.g., Lundgren et al. 2010). Accordingly, admitting $T_0 = 1.95 \; K$ as the upper bound of the DM temperature at the present epoch, we find that, the corresponding lower bound at transition is $0.322 \; K < T_{tr} \leq 0.369 \; K$. 

By virtue of Eq. (100), we can use Eq. (10), to determine (also) the variation of entropy, along a polytropic change of the DM fluid, from the epoch of transition to the present epoch. We obtain \be 1.666 \: {\cal C} \leq \left. \Dl {\cal S} \right \vert_{z_{tr} \rarrow 0} = {\cal S}_0 - {\cal S}_{tr} < 1.802 \: {\cal C} \: , \ee where ${\cal C}$ is given by Eq. (97). 

Finally, in view of Eq. (87), the combination of Eqs. (52) and (72) yields \be \frac{\varep_{int}}{\varep_{mat}} = \frac{1}{2 + 3 \vert \Gm \vert} \left ( \frac{1 + z_{tr}}{1 + z} \right )^{3(1 + \vert \Gm \vert)} , \ee which, at $z = z_{tr}$, results in \be \left. \frac{\varep_{int}}{\varep_{mat }} \right \vert_{tr} = \frac{1}{2 + 3 \vert \Gm \vert} \: . \ee We note that, in contrast to the common perception, in a polytropic-DM model, the onset of transition from deceleration to acceleration does not necessarily require $\varep_{int} > \varep_{mat}$. In fact, according to Eq. (52), equality between the internal (dark) energy density and its rest-mass counterpart occurred quite later, at $z = 0.384$ (for $\Gm = 0$), which is in good agreement with the corresponding observational ($\Lm$CDM) result, namely, $z = 0.391 \pm 0.033$ (Suzuki et al. 2012). Depending on $\Gm$ (in the range given by Eq. (94)), Eq. (103) suggests that, the transition from deceleration to acceleration takes place when \be 0.441 < \frac{\varep_{int }}{\varep_{mat}} \leq 0.500 \: . \ee The question is why it happens so. The answer is both revealing and simple: Because of the GR itself! 

\subsection{Why and when the Universe transits to acceleration}

In the context of GR, the dynamics of a homogeneous and isotropic, spatially-flat cosmological model, such as the one given by Eq. (22), is completely determined by Eqs. (30) and (40). Combining these two equations together, we obtain \be \frac{\ddot{S}}{S} = - \frac{4 \pi G}{3 c^2} \left ( \eps + 3 p \right ) \ee (see, e.g., Linder 2008; Caldwell \& Kamionkowski 2009) and, hence, the condition for accelerated expansion, $\ddot{S} > 0$, is, in fact, translated to \be \eps + 3 p < 0 \: . \ee Inserting Eqs. (3) and (39) of our (polytropic-DM) model into Eq. (106), and taking into account also Eqs. (33), (45), (51) and (87), we find that, the condition for the acceleration of the cosmological model (22) results in \bea && \varep + 3 p =  \rh_0 c^2 (1 + z)^3 \times \nn \\ && \left [ 1 - (2 + 3 \vert \Gm \vert) \frac{1 - \Om_M}{\Om_M} \frac{1}{(1 + z)^{3(1+ \vert \Gm \vert)}} \right ] < 0 \: ; \eea hence, a homogeneous and isotropic, spatially-flat cosmological model filled with matter in the form of a polytropic (DM) perfect fluid, most definitely accelerates its expansion, at cosmological redshifts lower than a specific value given by \be z < \left [ (2 + 3 \vert \Gm \vert) \frac{1 - \Om_M}{\Om_M} \right ]^{\frac{1}{3(1+ \vert \Gm \vert)}} - 1 \equiv z_{tr} \: , \ee in complete agreement to the transition redshift, $z_{tr}$, defined (in an independent manner) via Eq. (72). 

So, we conclude that, in order to determine the "why" and the "when" of the onset of cosmic acceleration, there may be no need for any exotic DE at all. Equivalently, a polytropic cosmic fluid could (most definitely) reveal such a reality, and, at the same time, it would illuminate the nature of the long sought DE, as due to the cosmic fluid's internal motions, whose catalytic role should be particularly emphasized.

\subsection{Additional constraints from CMB}

In order to tighten up the constraints on the various DE models, a common approach is to include additional information from the CMB, in the form of the so-called shift parameter, $R$, which is related to the position of the first acoustic peak in the power spectrum of the temperature anisotropies (Efstathiou \& Bond 1999). However, one should have in mind that, this parameter is not a directly-measured quantity. It is derived from the observational data only after assuming a specific model, usually the spatially-flat $\Lm$CDM model. Therefore, extra care is needed, when using $R$ to test more exotic DE models (see, e.g., Elgar{\o}y \& Multam{\"a}ki 2007).

The use of the shift parameter as a probe of DE is based on the observation that, different models will result in almost identical CMB power spectra (Efstathiou \& Bond 1999), if they have: \texttt{(i)} Identical CDM densities $\om_c = \Om_c h^2$ ($h$ is the dimensionless Hubble constant, defined by $H_0 = 100h \; km \: sec^{-1} \: Mpc^{-1}$), \texttt{(ii)} identical baryonic densities $\om_b = \Om_b h^2$, \texttt{(iii)} identical primordial fluctuation spectra, and \texttt{(iv)} identical values of the shift parameter, which, as far as spatially-flat models are concerned, is given by \be R = \sqrt{\Om_M} \int_0^{z_{*}} \frac{dz}{H(z)/H_0} \: , \ee where $z_{*}$ the value of the cosmological redshift at photon decoupling. 

In view of Eqs. (68) and (94), in our (polytropic-DM) cosmological model Eq. (109) is written in the form \be R = \int_0^{z_{*}} \frac{\left ( 1 + z^{\prime} \right )^{\frac{3}{2} \vert \Gm \vert} d z^{\prime}}{\left [ (1 - \Om_M) + \Om_M \left ( 1 + z^{\prime} \right )^{3 (1 + \vert \Gm \vert )} \right ]^{1/2}} \: . \ee Once again, the integral on the rhs of Eq. (110) can be solved explicitly in terms of hypergeometric functions (see e.g., Gradshteyn \& Ryzhik 2007 (7th Ed.), pp. 1005 - 1008), as follows \bea && R = \frac{2}{\left ( 2 + 3 \vert \Gm \vert \right ) \sqrt {1 - \Om_M}} \left [ (1 + z_{*})^{\frac{2 + 3 \vert \Gm \vert}{2}} \times \right . \nn \\ && \left . _2F_1 \left ( \frac{2 + 3 \vert \Gm \vert}{6 ( 1 + \vert \Gm \vert )}  \: , \: \frac{1}{2} \: ; \: \frac{8 + 9 \vert \Gm \vert}{6 ( 1 + \vert \Gm \vert )} \: ; \right . \right . \nn \\ && \left . - \left [ \frac{\Om_M}{1 - \Om_M} \right ] (1 + z_{*})^{3 (1 + \vert \Gm \vert )} \right ) - \nn \\ && \left . _2F_1 \left ( \frac{2 + 3 \vert \Gm \vert }{6 (1 + \vert \Gm \vert )}  \: , \: \frac{1}{2} \: ; \: \frac{8 + 9 \vert \Gm \vert }{6 ( 1 + \vert \Gm \vert )} \: ; \: - \left [ \frac{\Om_M}{1 - \Om_M} \right ] \right ) \right ] . \eea As far as the value of $z_{*}$ is concerned, we adopt the nine-year WMAP survey final result (Bennett et al. 2013), $z_{*} = 1091.64 \pm 0.47$, at 68\% confidence level ($CL$). Accordingly, for $\Gm = 0$, i.e., in the $\Lm$CDM-equivalent polytropic case, Eq. (111) yields \be R = 1.7342 \: , \ee while, according to the nine-year WMAP survey final results (Bennett et al. 2013), the value of the shift parameter in the standard $\Lm$CDM model is \be R = 1.7329 \pm 0.0058 \; (68\% \: CL) \: ; \ee in other words, the value of the shift parameter in the $\Lm$CDM limit of our (polytropic-DM) model reproduces, to high accuracy, the corresponding result obtained by fitting the observational (CMB) data to the standard $\Lm$CDM model. 

Although applicable only to $\Lm$CDM-like models, the nine-year WMAP survey final result given by Eq. (113) is compatible also to our $\Gm \neq 0$ model, up to $\vert \Gm \vert \leq 0.0271$ (Fig. 8). On the other hand, the range of values of $R$, arising from the combination of the Planck first-data release with those of the WMAP survey (Wang \& Wang 2013), namely, \be R = 1.7407 \pm 0.0091 \; (68\% \: CL) \: , \ee encapsulates the whole range of allowed values of $\Gm$ given by Eq. (94), yielding as the most preferable one the value $\vert \Gm \vert = 0.040$ (Fig. 8). Such a result conforms with the final conclusion of Wang \& Wang (2013) that, the addition the Planck distance priors to the current cosmological data, leads to a marginal inconsistency with a cosmological constant in a spatially-flat Universe.

\begin{figure}[ht!]
\centerline{\mbox {\epsfxsize=9.cm \epsfysize=7.cm
\epsfbox{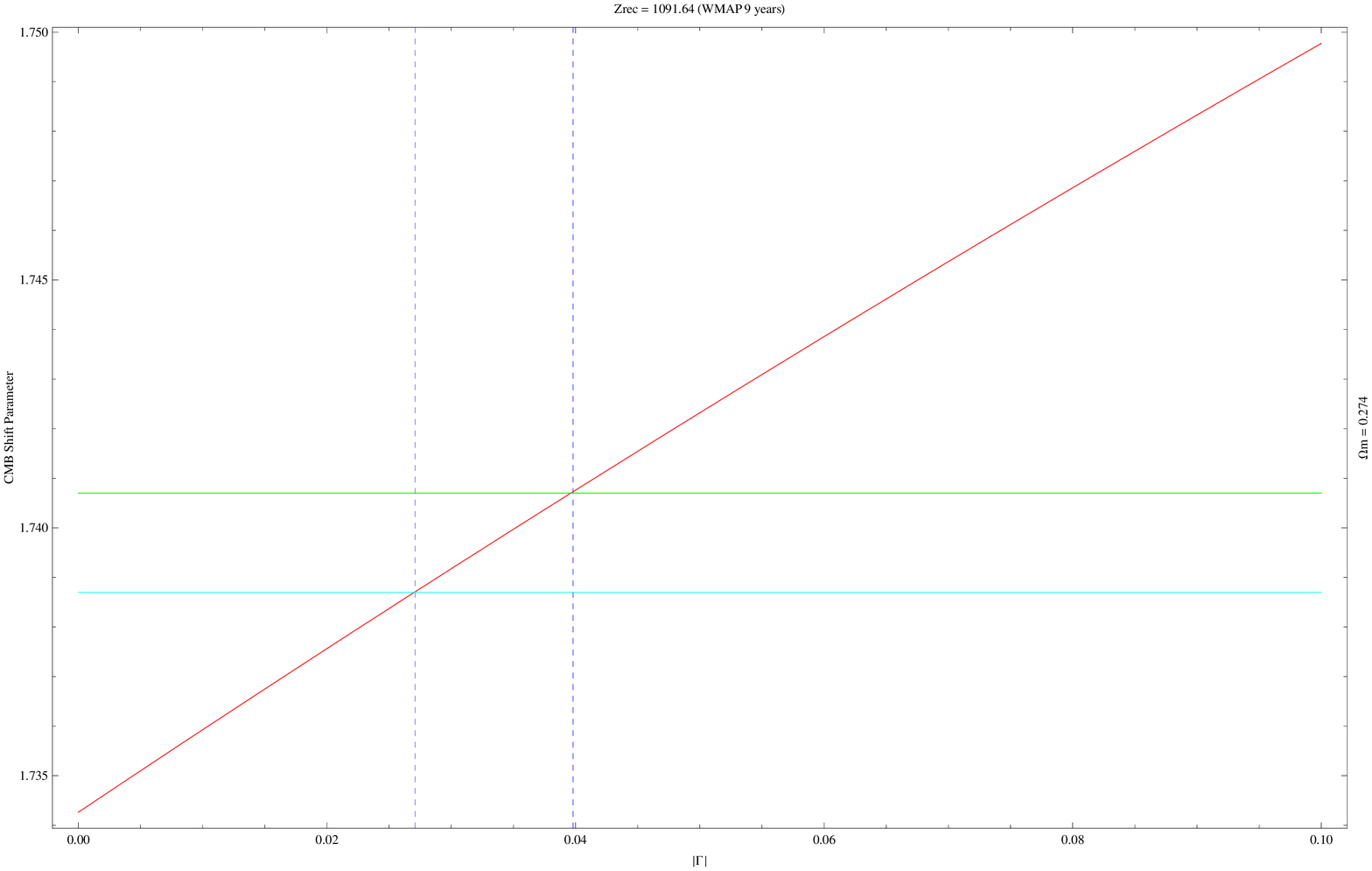}}} \caption{Theoretically (in the context of a polytropic-DM model) determined CMB-shift parameter, Eq. (111), as a function of the polytropic exponent $-0.089 < \Gm \leq 0$ (red solid line). The cyan horizontal straight line denotes the upper bound (at $68\% \; CL$) of the $R$-value, obtained from the nine-year WMAP survey final results, namely, $R = 1.7387$, and the corresponding green line denotes the mean value of $R$, arising from the combination of the Planck first-data release with those of the WMAP survey, namely, $R = 1.7407$. The vertical dashed lines denote, respectively, the upper bound of $\Gm$ $(\vert \Gm \vert = 0.0271)$, arising from WMAP alone, and the corresponding most preferable value $(\vert \Gm \vert = 0.0398)$, arising from the combination of the Planck first-data release with those of the WMAP survey.}
\end{figure}

In view of the aforementioned results, we conclude that, if someone wants to use the shift parameter as a constraint on a DE model, then, first of all, the distribution of the shift parameter has to be derived from the CMB data. This cannot be done without pre-assuming a model, i.e., without pre-determining the primordial power spectrum of density fluctuations (since, they form the basis for calculating the CMB anisotropies), and, therefore, making implicit assumptions about inflation. What is really worth noting, is that, the value of the CMB-shift parameter in the $\Lm$CDM limit of our (polytropic-DM) model (i.e., for $\Gm = 0$) reproduces, to high accuracy, the corresponding result obtained by fitting the observational (CMB) data to the standard $\Lm$CDM model. 

\section{Discussion}

In this article, we have examined the possibility that, the extra (dark) energy needed to flatten the Universe, is represented by the energy of the internal motions of a polytropic perfect fluid. Polytropic processes in a DM fluid have been most successfully used in modeling dark galactic haloes, improving significantly the velocity dispersion profiles of galaxies. Motivated by such results, we have explored the physical and the dynamical characteristics, as well as the evolution of a cosmological (toy-)model, driven by a gravitating fluid with (polytropic) thermodynamical content, consisting of DM (dominant) and baryonic matter (subdominant). 

In the distant past, the matter-energy content of this model behaves as a pressureless fluid (cf. Eq. (70), for $z \gg 1$). However, towards the present epoch, the internal physical characteristics of this fluid (i.e., beyond its rest-mass density) take over (cf. Eq. (20)), yielding the DM itself thermodynamically involved (in fact, at cosmological redshifts lower than $z = 0.384$, $\varep_{int} > \varep_{mat}$, i.e., the energy density of the internal motions dominates over the corresponding rest-mass quantity). 

In this context, the fundamental matter constituents of this model are the volume elements of the (DM) fluid, performing polytropic flows. As a consequence, the energy of this fluid's internal motions has also been taken into account as a source of the universal gravitational field. In this way, we have been able to determine the appropriate form of the cosmic scale factor (see Fig. 3), which (under the assumption that the DM is thermodynamically involved) governs the evolution of the Universe, being modeled as a spatially-flat RW spacetime. Accordingly, we have asked ourselves, whether this model can accommodate both the observed distribution of the cosmologically-distant indicators and the associated phase of accelerated expansion.

Our findings are, in fact, quite promising. In principle, the energy of the internal motions of the polytropic-DM fluid can account for the (extra) DE, so that, at the present epoch, the total-energy density parameter, $\Om = \frac{\varep}{\varep_{c}}$, equals to unity (cf. Eq. (50)). Furthermore, for values of the polytropic exponent, $\Gm$, lower than $0.541$, the (conventional) pressure of the (DM) fluid becomes negative enough (cf. Eq. (45)), in the sense that, the Universe accelerates its expansion at cosmological redshifts smaller than a transition value (cf. Eq. (72)), in a way consistent (also) with the requirement $\varep + 3p < 0$ (cf. Eqs. (107) and (108)). 

It is both pedagogic and quite interesting, to stress the various reasons for imposing successive constraints on the exact value of the polytropic exponent, $\Gm$. More specifically,

\begin{itemize}

\item the second law of thermodynamics in an expanding Universe suggests that $\Gm \leq \gm$ (cf. Eq. (13). In this context,

\item for the pressure work to be negative (i.e., to be attributed to the cosmic fluid itself), $\Gm < 1$ (cf. Eq. (20)). Furthermore,

\item the condition $z_{tr} \geq 0$ suggests that $\Gm \leq 0.541$ (cf. Eq. (74)), while,

\item the condition $p_0 > - \varep_0$ (i.e., the requirement of a non-phantom Universe) implies that $\Gm > - 0.377$ (cf. Eq. (55)). Moreover, 

\item the positivity of the velocity-of-sound square at all $z$, imposes the major constraint $\Gm \leq 0$ (cf. Eqs. (83) and (87)), and, finally,

\item the requirement for CDM at the present epoch implies that $\Gm > -0.089$ (cf. Eq. (93)).

\end{itemize}

In view of the aforementioned results, a realistic polytropic (DM) cosmological model (i.e., one that is compatible to the fundamental physical laws and the basic mathematical principles, and, at the same time, it is compatible to the observational data currently available) requires that, eventually, the polytropic exponent, $\Gm$, should be settled down to the range $-0.089 < \Gm \leq 0$, namely, if it is not zero, it is definitely negative and very close to zero (in connection, see Fig. 6). In any case, the polytropic-DM model attributes a well-posed physical meaning to $\Gm$ itself, suggesting that, its absolute value is, in fact, a representation of the value of $\left ( \frac{c_s}{c} \right )^2$ at transition (cf. Eq. (89)). 

In the context of the polytropic (DM) model, the ($\Gm$-dependent) present-time value of the deceleration parameter (Eq. (96)), is in good agreement with the observationally-determined range of values of $q_0$, namely, $q_0 = -0.53_{-0.13}^{+0.17}$ (Giostri et al 2012), while, for $\Gm = 0$, the equality between the internal (dark) energy density and its rest-mass counterpart occurs at $z = 0.384$ (cf. Eq. (52)), a theoretical prediction that (also) reproduces quite accurately the corresponding observational ($\Lm$CDM) result, namely, $z = 0.391 \pm 0.033$ (Suzuki et al. 2012).

Moreover, our cosmological model does not suffer either from the age problem (see Fig. 2) or from the coincidence problem (cf. Eqs. (52) and (53)), and, at the same time, it reproduces (to high accuracy) the distance measurements, performed with the aid of the SNe Ia standard candles (see Fig. 6). It is also worth noting that, the value of the CMB-shift parameter in the $\Lm$CDM limit of our (polytropic-DM) model (i.e., for $\Gm = 0$) reproduces, to high accuracy the corresponding value obtained by fitting the observational (CMB) data to the standard $\Lm$CDM model (cf. Eqs. (112) and (113)). Finally, the polytropic-DM model most naturally interprets not only when, but, also, why the Universe transits from deceleration to acceleration (cf. Eqs. (107) and (108)), and reveals the catalytic role of the internal motions in such a transition, i.e., for the onset of acceleration, the internal-energy density suffices to be half the value of the corresponding rest-mass quantity (cf. Eq. (104)). 

Nevertheless, there are (also) several aspects of the polytropic-DM model, that still remain open, such as:

\begin{itemize}

\item The cosmological behavior of the corresponding density perturbations: Since we do not neglect the pressure with respect to the energy density, the perturbations should be studied in a general-relativistic way. In this case, a problem arises on the choice of gauge. As it has already been recognized by Gondolo and Freese (2003), in a polytropic model, and (even) well-into the matter-dominated era, the value of the density perturbation depends on the choice of gauge. This creates a problem of interpretation for fluctuations in the present Universe, which must be addressed in future studies.

\item The origin of the (extra) amount of heat, ${\cal C} dT$, offered to the thermodynamical system (cf. Eq. (21)): Skillfully, this could be attributed to a long-range confining force between the DM constituents (see, e.g., Gondolo \& Freese 2003; Arkani-Hamed et al. 2009; Van den Aarssen et al. 2012), which, in our case, would be of the form $F = -Kr^{2 + 3 \vert \Gm \vert}$, where $r$ is the radial distance and $K > 0$ is a normalization constant (in connection, see Eq. (80) and the discussion following Eq. (89) of Gondolo and Freese (2003)). This force may be of gravitational origin or it may be a new force. However, it is not clear at all that, a system subject to a long-range confining force can reach thermodynamic equilibrium, hence, this matter must also be addressed in future studies.  

\end{itemize}

In any case, the assumption that the Universe matter content (basically its DM component) can be collisional (in the sense that it also possesses some sort of thermodynamical content), is to be seen as a (necessary) natural effort, to take into account all the (so far, practically, neglected) internal physical characteristics of a classical cosmological fluid, as sources of the universal gravitational field. Although speculative, the idea that the DE (needed to flatten the Universe) could be attributed to the internal motions of a polytropic-DM fluid, is (at least) intriguing and should be further explored and scrutinized in the search for conventional alternatives to the DE concept. 

\begin{acknowledgements}

The authors would like to thank Dr. Spyros Basilakos, for illuminating discussions and his useful comments on the content of this article. Financial support by the Research Committe of the Technological Education Institute of Central Macedonia at Serres, Greece, under grant SAT/ME/230113-10/04, is gratefully acknowledged. 

\end{acknowledgements}

\end{document}